%% file: main.tex
\documentclass[a4paper]{article}
\usepackage{times}
\usepackage{authblk}
\usepackage[]{algorithm}
\usepackage{algorithmicx}
\usepackage{algpseudocode}%
\usepackage{graphicx}
\usepackage{amsmath, amssymb}
\usepackage{booktabs}
\usepackage{color}
\usepackage{tikz}
\usetikzlibrary{positioning}
\usepackage[usenames,dvipsnames]{xcolor}
\usepackage[round]{natbib}
\usepackage{hyperref}

\newlength\myindent 
\usepackage[]{graphicx}
\chardef\bslash=`\\ 

\begin{document}

\title{A unified framework for spatially resolved cortical activation analysis}
\author[$1$, $2$*]{Lars Knieper}
\author[$3$]{Nadia Müller-Voggel}
\author[$4$]{Tobias Hepp}
\author[$1$]{Anna von Plessen}
\author[$1$, $2$]{Elisabeth Bergherr}

\affil[$1$]{ Chair of Spatial Data Science and Statistical Learning, University of G\"ottingen}
\affil[$2$]{ Campus Institute Data Science, University of G\"ottingen}
\affil[$3$]{ Departments of Neurosurgery, Universitätsklinikum Erlangen, Friedrich-Alexander University Erlangen-Nürnberg}
\affil[$4$]{ Institute of Medical Informatics, Biometry and Epidemiology, Friedrich-Alexander-Universität Erlangen-Nürnberg}
{\footnotesize{
\affil[*]{Corresponding author {\sf{e-mail: lars.knieper@uni-goettingen.de}}}}}
\maketitle  

\begin{abstract}
Cluster-based permutation tests are widely used for analyzing MEG data, even though they are limited to cluster-level inference and do not provide spatially resolved effect estimates.

We propose a regression-based framework for modeling brain activity directly on the cortical surface. Spatial effects are represented using Wendland radial basis functions, and model-based gradient boosting is employed for data-driven selection and estimation of localized activation patterns. This yields interpretable, spatially resolved effect estimates while mitigating the need for extensive multiple testing correction.
In a simulation study with heterogeneous signal structures, the proposed approach recovers localized effects that are difficult to detect using cluster-based methods. An application to experimental MEG data further illustrates its ability to reveal spatially specific activation patterns.

Overall, this unified framework provides a flexible and interpretable approach for modeling cortical surface data within a unified statistical setting.

\end{abstract}

\input{01introduction}

\input{02methods}

\input{03simus}

\input{04data}

\input{05conclusion}

\bibliographystyle{apalike}

\bibliography{bib}

\clearpage

\appendix
\input{06appendix}

\end{document}

%% file: 01introduction.tex
\section{Introduction}

Neuroimaging data, such as Magnetoencephalography (MEG) and electroencephalography (EEG), provide non-invasive measurements of brain activity with high temporal resolution. A common and recommended approach is to analyze these data in source space, where neural activity is projected onto the cortical surface \citep{gross2013goodpractice}. In this representation, brain measurements are defined on a high-dimensional and spatially structured domain, giving rise to the central scientific question of identifying where brain activity differs between experimental conditions, typically addressed through a contrast between condition-specific measurements for each subject.

A widely used approach to address this question is cluster-based permutation tests \citep{maris2007nonparametric}, which have become a standard tool in neuroimaging applications (see, e.g., \citealp{popov2017fef, oswald2017spontaneous, weisz2020auditory, vanes2019stimulus}). These methods aggregate neighboring vertices exceeding a predefined threshold into clusters and assess their significance via permutation-based inference. Their popularity stems from their ability to control the family-wise error rate while accounting for spatial dependence in the data.

However, cluster-based permutation tests are inherently limited to cluster-level inference and do not provide spatially resolved effect estimates. As pointed out by \citet{sassenhagen2019cptproblems}, precise conclusions about the spatial or temporal localization of effects are not supported, despite often being interpreted as such in practice. Moreover, no estimates of effect size or direction are provided, limiting the ability to compare activation patterns across regions or conditions. Recent work by \citet{rousselet2025badCPT} further emphasizes these limitations, while also noting favorable properties with respect to bias and variability. Nevertheless, the lack of effect estimation remains a fundamental restriction.

More generally, cluster-based permutation tests are not embedded in a regression framework. As a consequence, they do not naturally accommodate multiple covariates, subject-level predictors, or more complex modeling structures. In addition, aggregation within clusters may mask heterogeneous activation patterns, for instance when neighboring regions exhibit effects of differing magnitudes or even opposing signs.

Beyond cluster-based permutation tests, other work has applied spatially flexible or regression-based models to cortical or manifold-valued neuroimaging data, though typically targeting a different response type or scientific question than the one considered here.
In particular for component-wise gradient boosting, \citep{stocker2023functional} introduced an extension to manifold-valued outcomes, demonstrating its viability for shape and form regression; the response geometry there, however, is a shape/form quotient space rather than a fixed cortical sheet. \citet{ventrucci2014quasi} model resting-state quasi-periodic oscillations by fitting a small number of parametric dipole components on the flattened 2D sensor helmet, rather than estimating a spatially flexible field over the cortical surface itself. \citet{lila2016smooth} apply smooth functional principal component analysis with Laplace--Beltrami-based smoothing directly on the cortical surface, but target an unsupervised question -- the dominant modes of between-subject variation -- rather than an experimentally induced effect. Closest in spirit, \citet{Mejia02042020} propose a Bayesian spatial GLM for cortical fMRI measurements via an SPDE prior \citep{LINDGREN2022100599}, with activation regions subsequently identified through a separate, post-hoc excursion-set procedure rather than a jointly estimated selection mechanism. Accordingly, this pipeline is not intended for the same data situation, such that a full comparison would not be meaningful. An adapted version of this estimation and selection routine is nonetheless applied to provide additional simulation results and the data analysis in Appendix~\ref{app:bayesglm}.

Taken together, these limitations -- the cluster-level restriction of permutation-based inference and the narrower or fMRI-specific scope of existing spatial regression approaches -- highlight the need for methodological approaches that provide spatially resolved effect estimation, allow for flexible model extensions, and enable interpretability at the level of cortical regions. In particular, a regression-based framework facilitates the inclusion of additional covariates and supports a more detailed representation of the data-generating process.

To address these challenges, the analysis of cortical surface data is formulated as a functional regression problem \citep[overview in][]{morris2015functional}. Spatial activation effects are represented using localized Wendland radial basis functions \citep{wendland1995wendland} defined on the cortical surface, enabling flexible yet spatially interpretable modeling of spatial structure. Relevant basis functions are selected via component-wise gradient boosting (for an overview, see \citeauthor{Mayr2014EvolutionBoost}, \citeyear{Mayr2014EvolutionBoost}; for separate basis function selection, see short paper \citeauthor{hepp2025sparseboost}, \citeyear{hepp2025sparseboost}), resulting in sparse and localized effect estimates. 

This framework directly addresses the aforementioned limitations by providing spatially resolved effect estimation, supporting the inclusion of covariates within a regression setting, and allowing for heterogeneous effect patterns across neighboring regions. In contrast to cluster-based permutation tests, it enables both effect size estimation and interpretable localization within a unified statistical model. While the present work focuses on spatial effects, the modular structure readily allows extensions to incorporate temporal dynamics in future work.

To evaluate the suggested method, a comprehensive simulation framework is developed that generates realistic cortical activation patterns with heterogeneous signal structures and spatially correlated noise. The framework explicitly incorporates varying signal configurations, including differences in magnitude, sign, and spatial extent, allowing for a systematic assessment of method performance under controlled yet realistic conditions. 

Results demonstrate that the proposed approach reliably recovers localized and heterogeneous effects that may be difficult to detect using cluster-based methods. An application to experimental MEG data further illustrates its ability to reveal more nuanced spatial activation patterns.

The remainder of the paper is structured as follows. Section~\ref{sec:methods} introduces the methodological framework, including the functional regression formulation, spatial basis construction, and boosting procedure. Section~\ref{sec:sims} first lines out the simulation approach and then presents the results, and Section~\ref{sec:data} illustrates the approach on experimental MEG data. Section~\ref{sec:conclusion} concludes with a discussion of the results and future directions.

%% file: 02methods.tex
\section{Methods}
\label{sec:methods}

Since cluster-based permutation tests are widely used for the analysis of MEG and EEG data, they form the methodological benchmark for the approach proposed in this work. Their core idea is briefly outlined first, before the proposed regression-based framework for estimating spatially localized activation effects on the cortical surface is described in detail.

Throughout this work, spatial brain measurements are defined on the cortical surface mesh obtained from source reconstruction. The mesh consists of a set of discrete spatial locations, referred to as \textit{vertices}, which represent points on the cortical surface connected by edges forming triangular elements. Brain measurements are therefore observed at these vertices, and spatial relationships between them are defined by the mesh geometry.

All spatial distances used in the analysis are measured as geodesic distances along the cortical surface mesh in order to respect the intrinsic geometry of the cortex. Geodesic distances between two vertices are defined as shortest-path distances along the mesh graph, where edge weights correspond to Euclidean distances between adjacent vertices. A more detailed description of the geodesic distance computation is provided in Appendix~\ref{sec:geodesic}.

\subsection{Cluster permutation test}

Cluster-based permutation tests \citep{maris2007nonparametric}, as commonly implemented in the Python MNE library  \citep{gramfort2013MNE} or Matlab FieldTrip software \citep{oostenveld2011fieldtrip}, aim at exploring the spatial dependence of brain signals by aggregating adjacent locations that exceed a predefined test-statistic threshold into clusters. Statistical significance is assessed by comparing the observed cluster-level statistic to a permutation-based null distribution.

More precisely, a univariate test statistic (typically a $t$-statistic) is computed independently for each vertex (or originally each sensor). Vertices exceeding a predefined threshold corresponding to a chosen significance level are identified and spatially adjacent significant vertices are grouped into clusters. Then, a cluster-level statistic is calculated by summing up the test statistics within each cluster. The particular choice of the sum comes from the assumption of spatially extended, smooth effects, where signal accumulates across neighboring vertices. The significance of these observed clusters is evaluated by repeatedly permuting sign-flips on the vertex-wise measurements, and deriving permuted test-statistics and a null distribution of maximal cluster-level statistics across permutations.

This procedure has become a standard tool for mass-univariate analyses of brain activation data in particular because it controls the family-wise error rate at cluster level \citep{maris2007nonparametric}. This error rate refers to the probability to conclude if there is an effect due to the experimental condition even though there is not. A consequence is a particular sensitivity to spatially extended effects \citep{rousselet2025badCPT}.

However, inference done via cluster permutation tests is restricted to the cluster level. The method does not yield spatially resolved effect estimates, predictions, or a regression-based representation of the data-generating process. In particular, cluster permutation tests are not designed to jointly model multiple covariates or experimental factors, nor to estimate effect sizes in a way that allows direct comparison across conditions.

\subsection{Gradient boosted function-on-surface regression}

While cluster-based permutation tests address the question of whether an effect exists somewhere on the cortical surface, the approach proposed in this work focuses on estimating and predicting spatially sparse effect sizes within a regression framework.

To this end, preprocessed MEG source-space data are embedded into a functional regression framework that allows for a structured spatial representation of cortical activity. Rather than treating spatial locations independently, the cortical surface is represented by a two-dimensional manifold, which is subsequently partitioned into spatial basis functions. These basis functions serve as model components whose contributions are estimated and selected using model-based gradient boosting.

\begin{figure}[ht]
\centering
\begin{tikzpicture}[
    node distance=1cm,
    every node/.style={
        draw,
        rectangle,
        rounded corners,
        align=center,
        minimum height=1cm,
        minimum width=3.2cm
    },
    arrow/.style={
        ->,
        thick
    }
]

\node (preproc) {Source-space\\ cortical data};
\node (source) [right=of preproc] {Functional regression\\ framework\\ Section \ref{sec:functionregression}};
\node (basis) [below=of source] {Wendland radial\\ basis construction\\ Section \ref{sec:BF}};
\node (boost) [below=of basis] {Model-based\\ gradient boosting\\ Section \ref{sec:gradboost}};

\draw[arrow] (preproc) -- (source);
\draw[arrow] (source) -- (basis);
\draw[arrow] (basis) -- (boost);

\end{tikzpicture}
\caption{Pipeline of MEG data from preprocessing to the spatial effect estimation using model-based gradient boosting.} \label{fig:analysis_pipeline}
\end{figure}
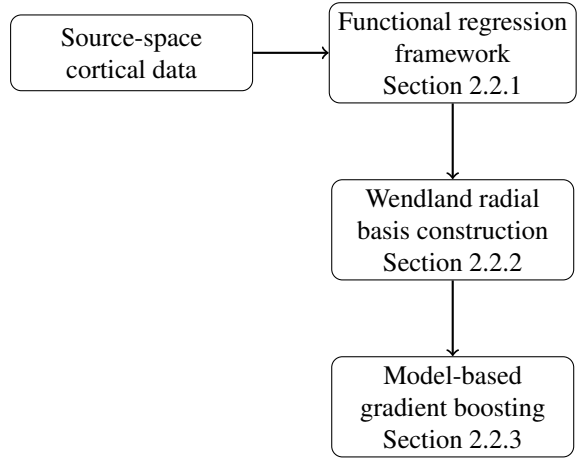

Figure~\ref{fig:analysis_pipeline} illustrates the proposed analysis pipeline, which will be described in more detail in the subsequent sections. The explanation focuses on the statistical methodological parts starting with an embedding of source-reconstructed data into a functional regression framework. It follows with a detailed explanation of how the effects on the cortical surface are modeled with Wendland radial basis functions, which effects are estimated and selected via model-based gradient boosting.

\subsubsection{Functional regression framework}
\label{sec:functionregression}

Each individual brain is viewed as a single functional observation defined over the cortical surface. Specifically, for subject $i = 1, \dots, N$, let 
$\boldsymbol{Y}_i(\boldsymbol{s})$
denote the observed brain measurements at cortical vertices $v = 1, \dots, n_v$ with three-dimensional spatial coordinates $\boldsymbol{s}_v = (s_{vx}, s_{vy}, s_{vz})^\top$.
Thus, the number of observations corresponds to the number of subjects, while each observation is a function of dimension $(n_v \times 1)$ evaluated on a cortical surface. A brain of subject $i$ is therefore modeled as
\begin{equation}
\label{eq:brain_mod}
\boldsymbol{Y}_i(\boldsymbol{s}) = f(\boldsymbol{s}) +  \boldsymbol{\varepsilon}_i(\boldsymbol{s}),
\end{equation}
where $f(\boldsymbol{s})$ represents a spatial activation field associated with the experimental condition and $\boldsymbol{\varepsilon}_i(\boldsymbol{s})$ denotes subject-specific noise. The latter is assumed to be spatially correlated within each observation $i$. Such spatial correlations arise due to task-irrelevant cognitive processes, measurement noise, and the spatial smoothness induced by forward modeling and preprocessing steps. Consequently, $\boldsymbol{\varepsilon}_i(\boldsymbol{s})$ captures structured variability that is not caused by the experimental manipulation and varies across subjects. 
From a functional regression perspective, model \eqref{eq:brain_mod} may be interpreted as an intercept-only function-on-surface regression model, where $f(\boldsymbol{s})$ acts as a smooth functional intercept. This formulation enables straightforward extensions in future work.

The primary goal is to identify spatial regions of the cortex that exhibit activation related to the experimental condition. These regions are encoded in the spatial activation function $f(\boldsymbol{s})$, which is assumed to be shared across individuals. This assumption reflects that task-related activations occur at similar cortical locations across subjects, but actual measurements may differ.

Modeling $f(\boldsymbol{s})$ comes with challenges. First, brain measurements are strongly spatially correlated with the number of vertices typically exceeding the number of subjects substantially, $n_v \gg N$.
Second, activation effects are expected to be spatially localized rather than globally smooth. Common spatial regression modeling techniques typically operate globally over the whole spatial domain and do not naturally provide a mechanism for identifying localized activation regions. This motivates a modeling strategy that allows not only for a flexible representation of the spatial activation but also for an identification of which spatial regions contribute to the activation pattern.

In the following subsections, we describe how spatial basis functions are constructed along the cortical surface and get selected using model-based gradient boosting.

\subsubsection{Wendland radial basis functions}
\label{sec:BF}

Equation~(\ref{eq:brain_mod}) introduces an unknown spatial activation function $f(\boldsymbol{s})$ defined on the cortical surface. The functional form of this effect is unspecified, but by assumption the condition-related brain activation varies smoothly across neighboring locations while remaining spatially localized. Consequently, a flexible representation is required that allows for smooth spatial effects while preserving local interpretability.

A common strategy for representing smooth spatial functions is to use basis function expansions within the framework of generalized additive models (GAMs; \citealp{hastie1986gams}). In this work, the spatial
activation function $f(\boldsymbol{s})$ is approximated using Wendland radial basis functions (WRBF; \citealp{wendland1995wendland}). These basis functions are compactly supported and therefore locally defined on the spatial domain. This property is particularly advantageous in the present setting, as it allows spatial effects to be represented by localized model components that can be selected individually during model fitting. In contrast, globally supported smoothers such as thin plate splines \citep{wood2003tps}, which are more common, are influenced across the entire spatial domain and therefore do not naturally facilitate localized effect selection.

Let $d(\boldsymbol{s},\boldsymbol{s}')$ denote the geodesic distance between two surface locations, computed as the shortest-path distance along the cortical mesh graph with edge weights given by Euclidean distances between neighboring vertices. To construct the basis functions, $J$ center locations $\boldsymbol{c}_1,\dots,\boldsymbol{c}_J$ are selected on the cortical surface using geodesic farthest-point sampling, which ensures an approximately uniform coverage of the spatial domain. For a fixed radius parameter $r > 0$, the $j$-th radial basis function is defined as

$$
b_j(\boldsymbol{s})
=
\phi\!\left(
\frac{d(\boldsymbol{s}, \boldsymbol{c}_j)}{r}
\right),
\qquad j = 1,\dots,J,
$$

where $\phi(\cdot)$ denotes a compactly supported radial kernel. In the present work we employ the Wendland $C^2$ kernel

$$
\phi(u) = (1-u)^4(4u+1) \cdot \mathbf{1}(u < 1),
$$

which yields basis functions that smoothly decrease to zero at the boundary of their support. The parameter $r$ controls the spatial extent of each basis function and therefore determines the degree of locality of the representation.

\begin{figure}[htbp]
\centering

\begin{minipage}{0.48\textwidth}
  \centering
  \includegraphics[trim={3cm 3cm 1cm 0.3cm},clip,width=\linewidth]{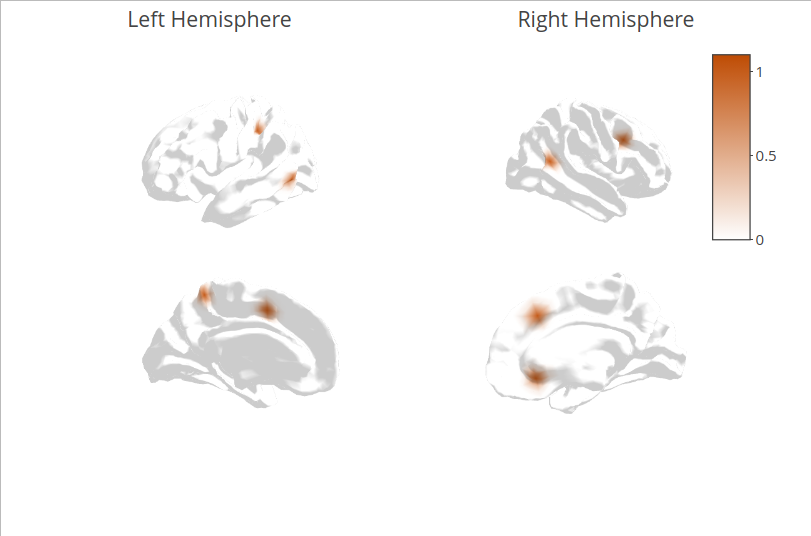}
\end{minipage}\hfill
\begin{minipage}{0.48\textwidth}
  \centering
  \includegraphics[trim={3cm 3cm 1cm 0.3cm},clip,width=\linewidth]{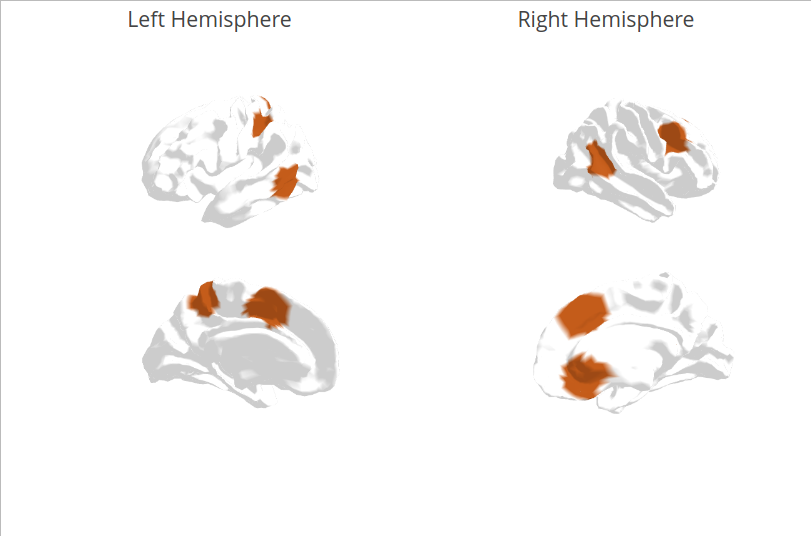}
\end{minipage}
\caption{Examples of Wendland radial basis functions centered at different cortical locations. Left panel: Unscaled basis function evaluations; right panel: binary representation of regions covered by these particular basis functions.} \label{fig:basis_fcts}
\end{figure}

Throughout simulations and the data application radius of $r = 0.02$ is applied such that one basis function covers $68.05$ vertices on average\footnote{This particular choice showed to yield a smooth estimation as well as as a good selection of space.}. Figure~\ref{fig:basis_fcts} illustrates unscaled basis function evaluations on the left panel while the corresponding covered regions in a binary representation on the right panel. The number of basis functions needs to be sufficiently large such that basis functions overlap and therefore account for the neighborhood structure. Throughout this work we apply $J =500$ functions at different centers $c_j$, which leads to the fact that one vertex is covered by $8.31$ basis functions on average.

Using these basis functions, the unknown activation field can be expressed as a linear expansion

$$
f(\boldsymbol{s}) =
\sum_{j=1}^{J}
\theta_j \, b_j(\boldsymbol{s}),
$$

where $\theta_j$ denotes the coefficient associated with the $j$-th basis function.

Since cortical coordinates are identical across subjects, the basis functions are shared across all observations. For a vertex $v=1,\dots,n_v$ with spatial coordinate $\boldsymbol{s}_v$, the spatial effect can therefore be written as
\[
f(\boldsymbol{s}_v) =
\boldsymbol{b}(\boldsymbol{s}_v)^\top
\boldsymbol{\theta},
\]
where $\boldsymbol{b}(\boldsymbol{s}_v) =
(b_1(\boldsymbol{s}_v),\dots,b_J(\boldsymbol{s}_v))^\top$ contains the WRBF evaluated at vertex $v$, and
$\boldsymbol{\theta} = (\theta_1,\dots,\theta_J)^\top$ contains the corresponding coefficients.

Stacking these vectors across all vertices yields the spatial design matrix
\[
\mathbf{X} =
\begin{pmatrix}
\boldsymbol{b}(\boldsymbol{s}_1)^\top \\
\vdots \\
\boldsymbol{b}(\boldsymbol{s}_{n_v})^\top
\end{pmatrix}
\in \mathbb{R}^{n_v \times J}.
\]

For each subject $i$, the functional regression model introduced in Equation~\ref{eq:brain_mod} can therefore be written as
\[
\boldsymbol{Y}_i(\boldsymbol{s}) =
\mathbf{X}\boldsymbol{\theta}
+
\boldsymbol{\varepsilon}_i(\boldsymbol{s}).
\]

Stacking the responses across all $N$ subjects results in the vectorized model
\[
\boldsymbol{Y}(\boldsymbol{s}) =
(\mathbf{1}_N \otimes \mathbf{X})\,
\boldsymbol{\theta}
+
\boldsymbol{\varepsilon}(\boldsymbol{s}),
\]
where $(\mathbf{1}_N \otimes \mathbf{X})$ replicates the spatial design matrix for each subject.

This representation expresses the spatial activation field as a linear combination of localized basis functions. Each basis function represents a spatially restricted region of the cortical surface and can therefore be interpreted as a candidate activation component. Treating the WRBF individually as model components forms the basis for their subsequent selection and estimation using model-based gradient boosting (Section~\ref{sec:gradboost}).

\subsubsection{Gradient boosting}
\label{sec:gradboost}

Component-wise model-based gradient boosting (\citealp{Buehlmann2003ComponentBoosting,Buehlmann2006BoostingHighDim}, overview in \citealp{Mayr2014EvolutionBoost}) is a flexible framework for estimation of statistical models while simultaneously performing variable selection. The component-wise structure allows it to handle cases where the number of parameters exceeds the number of observations and early stopping of the boosting iterations serves as a form of regularization \citep{Mayr2012Stopping}.
These properties make model-based gradient boosting particularly useful for settings with a large number of candidate model components/variables. In the present work, high dimensionality arises from the representation of the spatial activation field by WRBF, where $J = 500$ candidate basis functions will be used.

Within this framework, model-based gradient boosting provides a data-driven mechanism to determine whether a coefficient $\theta_j$ associated with a given basis function is effectively zero and therefore, suggests that no condition-related activation is present in the corresponding spatial region. Although the resulting spatial effect is represented as a smooth, non-linear function over the cortical surface, the individual basis functions enter the model linearly. This allows model-based gradient boosting to be applied using linear model components only, while still yielding a non-linear effect estimate.

Algorithm \ref{alg:boosting_algo} presents the resulting model-based gradient boosting procedure for WRBF. The algorithm requires the specification of a total number of boosting iterations $m_{\text{stop}}$, which needs to be chosen sufficiently large, as well as a step length, typically set to $\nu = 0.1$. The parameter vector is typically initialized as $\widehat{\boldsymbol{\theta}}^{[0]} = \boldsymbol{0}$ (line {\small1}). 

In each iteration $m$ the gradient of pre-specified loss function $\rho$ is computed. For normally distributed responses, the loss function corresponds to the squared error 
\begin{equation} \label{eq:lossfct}
    L (\boldsymbol{y}(s), \widehat{\boldsymbol{\theta}}) = 
    \left( \boldsymbol{Y}(\boldsymbol{s}) - (\mathbf{1}_N \otimes \mathbf X)\, \widehat{\boldsymbol{\theta}} \right)^2,
\end{equation}
such that the resulting gradient is equal to the residuals, that is, the discrepancy between the observed data and the fitted values given the available covariates. Consequently, the gradient at iteration $m$ is defined as
\begin{equation} \label{eq:gradient}
    \boldsymbol{u}^{[m]} = 
    \boldsymbol{Y}(\boldsymbol{s}) - (\mathbf{1}_N \otimes \mathbf X)\, \widehat{\boldsymbol{\theta}}^{[m-1]},
\end{equation}
where $\widehat{\boldsymbol{\theta}}^{[m-1]}$ denotes the coefficient estimates obtained in the previous iteration (line {\small3}).

These residuals serve as the target for refitting each WRBF separately in a linear regression fashion (line {\small5}). The basis function that explains the remaining residual variation best is selected (line {\small7}) and its contribution is added to the current coefficient vector gradually by step size $\nu$ (line {\small8}). Repeating this procedure for $m_{\text{stop}}$ iterations builds up a spatial activation field. The optimal stopping iteration $m^*$ is determined by a pre-specified criterion (line {\small10}), yielding the final regularized coefficient vector $\widehat{\boldsymbol{\theta}}^{[m*]}$ (line {\small11}).

\begin{algorithm}[H]
\caption{Model-based gradient boosting for Wendland radial basis functions}\label{alg:boosting_algo}
\begin{algorithmic}[1]
    \State \textbf{Initialize} number of total iterations $m_\mathrm{stop}$, step length $\nu$ and $\widehat{\boldsymbol{\theta}}^{[0]}$
    \For{$m = 1$ to $m_\mathrm{stop}$}
        \State Compute $\mathbf{u}^{[m]}$ as stated in Equation \ref{eq:gradient}
         \For{each basis function $j = 1, \ldots, J$}
            \State Fit $\mathbf{u}^{[m]}$ to single basis function:
            \Statex \hspace{4.3em} $\hat{\theta}^{[m]}_j = \arg \min_{\theta} \sum_{i = 1}^{N n_v} \left( u_i^{[m]} - \theta\, X_{ij} \right)^2$
        \EndFor
        \State Select basis function $j^*$ that fits $\mathbf{u}^{[m]}$ best:
        \Statex \hspace{2.6em} $j^* = \underset{j}{\arg \min} \sum_{i = 1}^{N n_v} \left( u_i^{[m]} - \hat{\theta}^{[m]}_j X_{ij} \right)^2$
        \State Update coefficient vector $\widehat{\boldsymbol{\theta}}^{[m]}$ at position $j^*$ only:
        \Statex \hspace{4.3em} $\widehat{\boldsymbol{\theta}}^{[m]}_{j^*} \leftarrow \widehat{\boldsymbol{\theta}}^{[m-1]}_{j^*} + \nu\, \hat{\theta}^{[m]}_{j^*}$

    \EndFor
    
    \State \textbf{Determine} optimal stopping iteration $m^*$ based on a chosen criterion
    \State \textbf{Return} final coefficient vector $\widehat{\boldsymbol{\theta}}^{[m*]}$
\end{algorithmic}
\end{algorithm}

\paragraph{Early Stopping}

In model-based gradient boosting, the number of boosting iterations is the most important tuning parameter, as it directly controls the degree of regularization and the number of selected model components \citep{Mayr2012Stopping}. A standard and widely used approach for determining the optimal stopping iteration $m^*$ is a $k$-fold cross-validation \citep{Stone1974CV}. However, component-wise boosting is known to result in rather rich models in terms of variable selection, such that several methodological extensions are suggested (see, e.g., \citealp{hofner2015stabsel, thomas2017probing, ellenbach2021robC, stromer2022deselection}).

In this work, $k$-fold cross-validation is applied at the subject level. Hence, the set of subjects is partitioned into $k$ disjoint folds. For each fold $l = 1, \dots, k$, the model is trained on $N_l = N (1 - \frac{1}{k})$ subjects and evaluated on the held-out $N_{-l} = \frac{N}{k}$ subjects. For each boosting iteration $m$, the out-of-sample loss (using Equation \ref{eq:lossfct}) is computed on the test subjects and averaged across folds. The optimal stopping iteration is then defined as

$$
m^* = \underset{m}{\text{arg min}} \; \; \frac{1}{k} \sum^k_{l=1} \frac{1}{N_{-l}} L\left(\boldsymbol{Y}_{-l}(\boldsymbol{s}), (\mathbf{1}_{N_{-l}} \otimes \mathbf X_{-l})\, \widehat{\boldsymbol{\theta}}^{[m]}\right).
$$

Algorithm \ref{alg:boosting_algo} is then run for $m^*$ iterations and the resulting coefficient vector $\widehat{\boldsymbol{\theta}}^{[m^*]}$ is taken as the final model. However, in high-dimensional settings with strongly correlated predictors - such as the present WRBF - this procedure may still select a relatively large number of components.

To further improve sparsity without increasing the overall degree of regularization, the deselection approach \citep{stromer2022deselection} is applied. After fitting the model up to iteration $m^*$, the cumulative loss reduction contributed by each selected component, denoted by $\Delta L_j$ , is computed and compared to the total loss reduction $\Delta L_{\text{tot}}$ achieved between the initial model $m^0$ and $m^*$. Given a threshold $\kappa \in (0, 1)$, all components satisfying $\Delta L_j < \kappa \; \Delta L_{\text{tot}}$ are removed from the model. The boosting procedure is then re-run using only the remaining components for $m^*$ iterations. In both simulation studies and data applications, a threshold of $\kappa = 0.01$ is used.

This deselection approach is particularly well suited to the present application, where neighboring basis functions and thus adjacent cortical regions are highly correlated. While standard gradient boosting might select many components, deselection focuses the final model on those basis functions that contribute most strongly to explaining the observed brain activation, thereby improving the interpretability of the estimated spatial effects. Therefore, a usual $k$-fold cross-validation determines the optimal stopping iteration, while deselection further sparsifies the resulting model by retaining only those basis functions that contribute most strongly to explaining the observed brain activation.

%% file: 03simus.tex
\section{Simulations}
\label{sec:sims}

The main goal of the simulation study is to generate realistically heterogeneous yet reproducible vertex-wise brain measurement fields for method evaluation. Simulations are conducted in a surface-based source space on the white-matter surface of the FreeSurfer \textit{fsaverage} brain (oct6 spacing) \citep{Fischl2012FreeSurf}. In each hemisphere, $n_v = 4098$ cortical vertices $v = 1, \dots, n_v$ are simulated.

\subsection{Setting}

Rather than generating brain measurements for two experimental conditions and computing their differences post hoc, simulations are performed directly on the level of conditional differences. Spatial \textit{signal plateau regions} $p = 1, \dots, P$ with signal amplitudes $\tau_p$ are defined, representing true activation of brain vertices induced by a hypothetical condition. To mimic spatial signal leakage arising from source reconstruction and spatial smoothing effects, smooth transition regions -- referred to as \textit{halos} -- are generated around each signal region. In contrast to the signal regions themselves, whose spatial support is fixed across all individuals, halo effects vary across subjects. Additionally, spatially correlated Gaussian process noise, denoted by $\boldsymbol{\varepsilon}$, is added to introduce heterogeneity across simulated brains. The resulting conditional-effect measurement at vertex $v$ is defined as

$$
y_v = \sum_{p=1}^{P} \tau_p w_{vp} + \varepsilon_v.
$$

The individual components of the simulation model are described in more detail below.

\paragraph{Signal plateau regions}

Signal plateau regions $p = 1, \dots, P$ are defined directly on the cortical surface using geodesic distances along the mesh. 
Let $d_{vp} = d(v, c_p)$ denote the geodesic distance between vertex $v$ and a manually chosen center vertex $c_p$ of plateau region $p$. 
Given a region-specific radius $r_p$, the set of vertices belonging to signal plateau region $p$ is defined as

\[
\mathcal{P}_p = \{ v : d_{vp} \le r_p \}.
\]

All vertices within $\mathcal{P}_p$ receive a constant signal amplitude $\tau_p$, producing a flat signal plateau on the cortical surface. 
The spatial support of the signal regions is identical for all simulated subjects, ensuring a fixed ground truth for evaluating vertex-wise detection performance.

\paragraph{Halos}

To recreate the spatial spread of localized neural activation to neighboring vertices induced by source reconstruction and spatial smoothing, and to introduce additional inter-individual variability in cortical measurements, smooth halo regions are generated around each signal plateau region. While the signal regions themselves remain identical across subjects, halo effects are allowed to vary between subjects by drawing region-specific decay parameters $\rho_p$ for each subject and each plateau region. Each $\rho_p$ is therefore uniformly distributed, $\rho_p \sim \mathcal{U}(0.015, 0.03)$.

Let $d_{vp}$ again denote the geodesic distance between vertex $v$ and the center of plateau region $p$. A halo decay is applied to vertices outside the signal region using a Gaussian kernel

$$
w_{vp} =
\begin{cases}
\exp\!\left(
-\dfrac{(d_{vp}-r_p)^2}{2\rho_p^2}
\right), & d_{vp} > r_p, \\
1, & d_{vp} \le r_p.
\end{cases}
$$

where $\rho_p$ controls the spatial decay of the halo. For vertices inside the signal region ($d_{vp} \le r_p$), the weight is defined as $w_{vp}=1$, ensuring a plateau of constant signal amplitude within the signal region. 

The resulting signal with halo contribution at vertex $v$ is therefore

$$
\eta_v = \sum_{p=1}^{P} \tau_p w_{vp}.
$$

Since two signal plateau regions are simulated per hemisphere, halo effects may spatially overlap, leading to vertices with combined signal strengths $\eta_v > \tau_p$. The resulting spatial field therefore consists of flat signal plateaus surrounded by smoothly decaying halo regions whose spatial extent varies between subjects.

\paragraph{Noise}

To account for spatially correlated measurement noise and inter-individual variability, a Gaussian random field is simulated on the cortical surface mesh.

Let $d(v,v')$ denote the geodesic distance between vertices $v$ and $v'$ along the cortical surface. A spatial covariance matrix $\mathbf{K} \in \mathbb{R}^{n_v \times n_v}$ is defined using a Gaussian kernel

$$
K_{vv'} =
\exp\!\left(
-\frac{d(v,v')^2}{2\ell^2}
\right),
$$

where $\ell$ controls the spatial correlation range. Spatial noise vectors $\boldsymbol{\varepsilon}$ are then drawn from the multivariate normal distribution

$$
\boldsymbol{\varepsilon} \sim
\mathcal{N}(\mathbf{0}, \sigma^2 \mathbf{K}),
$$

where $\sigma^2$ controls the marginal variance of the noise process. For computational efficiency, noise fields are generated using a spectral representation of the covariance matrix (for more details see Appendix~\ref{app:GaussianNoise}).

\begin{figure}[htbp]
\centering

\begin{minipage}{0.48\textwidth}
  \centering
  \includegraphics[trim={3cm 3cm 1cm 0.3cm},clip,width=\linewidth]{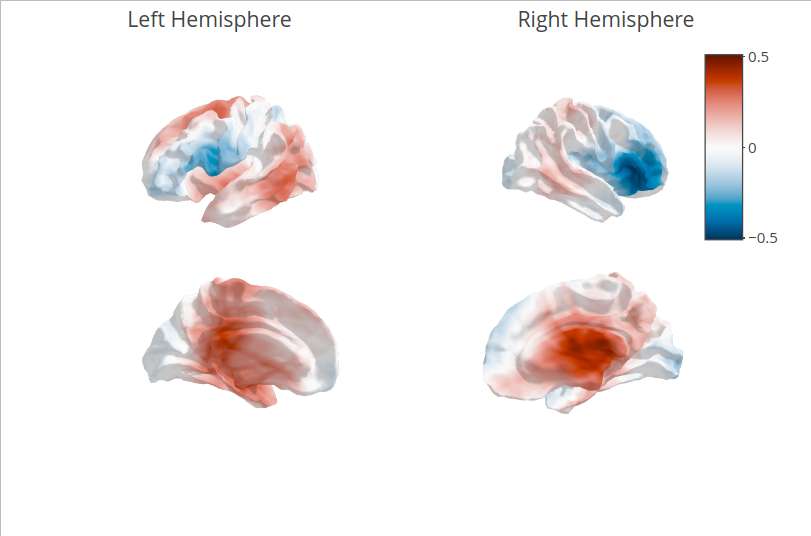}
\end{minipage}\hfill
\begin{minipage}{0.48\textwidth}
  \centering
  \includegraphics[trim={3cm 3cm 1cm 0.3cm},clip,width=\linewidth]{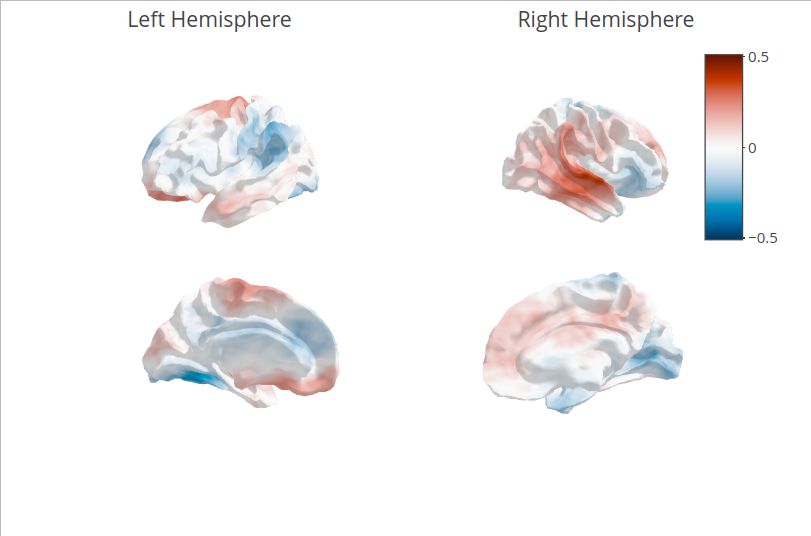}
\end{minipage}

\vspace{0.5em}

\begin{minipage}{0.48\textwidth}
  \centering
  \includegraphics[trim={3cm 3cm 1cm 0.3cm},clip,width=\linewidth]{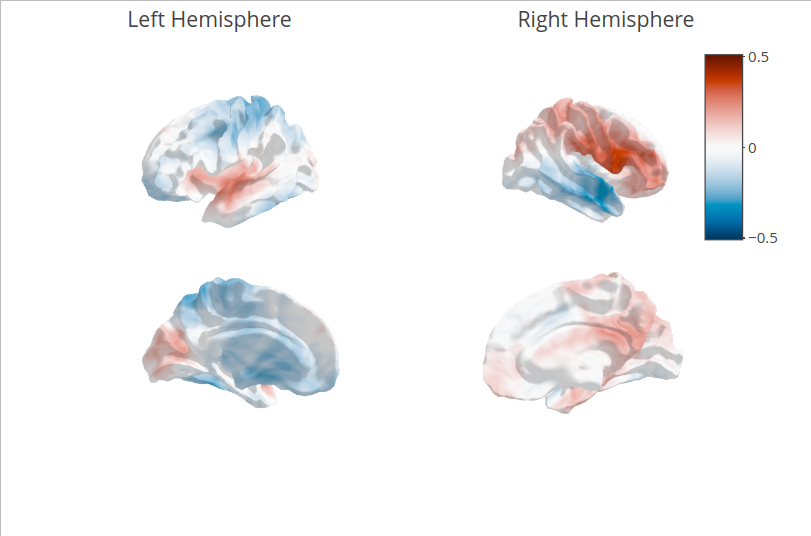}
\end{minipage}\hfill
\begin{minipage}{0.48\textwidth}
  \centering
  \includegraphics[trim={3cm 3cm 1cm 0.3cm},clip,width=\linewidth]{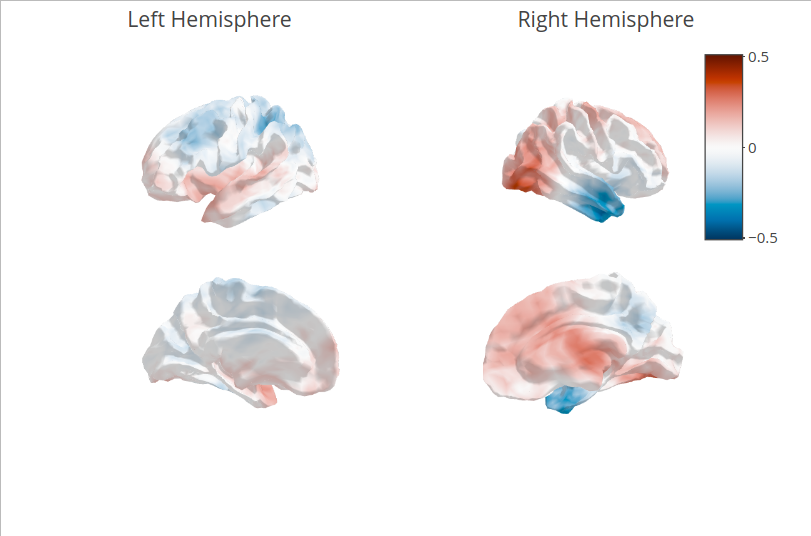}
\end{minipage}
\caption{Examples of simulated spatial Gaussian noise fields only for four distinct subjects (without signal or halo components).} \label{fig:sim_noise}
\end{figure}

For the simulation study setting $\sigma = 0.12$ and $\ell = 0.04$ results in heterogeneous spatial noise fields as illustrated in Figure~\ref{fig:sim_noise}. Due to the relatively large variance, noise can locally exceed the signal and its halo at the single-subject level. However, since the noise has mean zero, averaging across subjects the signal with its halo becomes increasingly visible (see Section~\ref{sec:sim_metrics_scenarios} for examples of averaged effects under different simulation scenarios).

This component-wise procedure of signal, halo and noise yields subject-specific spatial fields with fixed signal region support, heterogeneous halo effects surrounding these regions, and spatially correlated noise across the cortical surface. The chosen parameter settings are guided by empirical characteristics observed in real MEG data (see Section~\ref{sec:data}). Examples of complete simulated brain measurements are provided in Appendix~\ref{sec:App_sim_measurements}.

\subsection{Metrics and Scenarios}
\label{sec:sim_metrics_scenarios}

\paragraph{Metrics}

The simulation study is designed to assess whether the proposed methods are able to (i) detect spatially localized signal plateau regions, (ii) control the spatial extent of detected regions, and (iii) accurately estimate their effect sizes. These objectives are evaluated under varying signal configurations and sample sizes.

To assess the ability to correctly detect signal regions (i), a vertex-wise true positive rate ($TPR$) is considered. This is defined as the proportion of vertices within the true signal regions that are detected by the two methodological approaches. For the boosting approach, a vertex is counted as detected if it is covered by at least one selected basis function. For cluster-based permutation tests, a vertex is considered detected if it belongs to a significant cluster.

To evaluate whether detection is achieved without excessive spatial spread, a vertex-wise false positive rate ($FPR$) is considered. This is defined as the proportion of vertices outside the true signal regions that are nevertheless identified by a given method. Analogous to the definition of $TPR$, vertices are counted towards the $FPR$ if they are selected by at least one covering basis function in the boosting approach, or, in the case of cluster-based permutation tests, if they belong to a significant cluster despite lying outside the true signal plateau regions.

These two metrics allow for a direct comparison between both methods with respect to detection performance and spatial specificity.

To further evaluate the accuracy of effect estimation, mean absolute error and root mean squared error are computed between the estimated effect at each vertex, $\widehat{f(\boldsymbol{s})_v}$, and the true underlying signal amplitude $\tau_v$:

$$
\text{mae}_\tau = \frac{1}{n_v} \sum_{v=1}^{n_v} \left|\widehat{f(\boldsymbol{s}_v)} - \tau_v \right|, \quad
\text{rmse}_\tau =
\sqrt{\frac{1}{n_v} \sum_{v=1}^{n_v} \left( \widehat{f(\boldsymbol{s}_v)} - \tau_v \right)^2}.
$$

This comparison is performed with respect to the true plateau signals without halo contributions, thereby isolating recovery of the primary signal component. Since cluster-based permutation tests do not provide effect size estimates, these metrics are only available for the proposed boosting approach.

\paragraph{Simulation settings}

All metrics are evaluated under two simulation scenarios with varying signal configurations. As analyses are conducted separately for each hemisphere, each hemisphere represents a distinct setting with specific signal characteristics\footnote{Since both hemispheres share highly comparable cortical mesh structures, the assignment of signal configurations to the left or right hemisphere is arbitrary and does not influence the results.}. These settings are therefore named according to their signal structure:

\begin{itemize}
    \item \textbf{Varying signs (LH):} Two signal regions with equal magnitude but opposite signs: $\tau_1 = 0.15$ (radius $r_1 = 0.02$); $\tau_2 = -0.15$ ($r_2 = 0.015$).
    \item \textbf{Varying amplitudes (RH):} Two positive signal regions with differing amplitudes: $\tau_1 = 0.2$ (radius $r_1 = 0.02$); $\tau_2 = 0.15$ ($r_2 = 0.015$).
    \item \textbf{Varying signs and amplitudes (LH):} Two signal regions with opposing signs and unequal amplitudes:  $\tau_1 = 0.2$ (radius $r_1 = 0.02$); $\tau_2 = -0.15$ ($r_2 = 0.015$).
    \item \textbf{Single signal (RH):} One signal plateau with a second region absent:  $\tau_1 = 0.2$ (radius $r_1 = 0.02$); $\tau_2 = 0$.
\end{itemize}

\begin{figure}[t]
\centering
\begin{minipage}{0.48\textwidth}
  \centering
  \includegraphics[trim={4cm 4cm 1cm 0.3cm},clip,width=\linewidth]{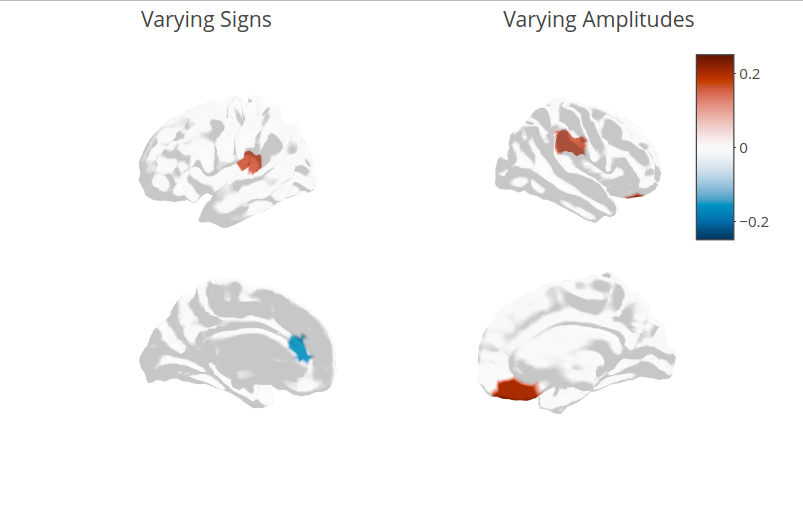}
\end{minipage}\hfill
\begin{minipage}{0.48\textwidth}
  \centering
  \includegraphics[trim={4cm 4cm 1cm 0.3cm},clip,width=\linewidth]{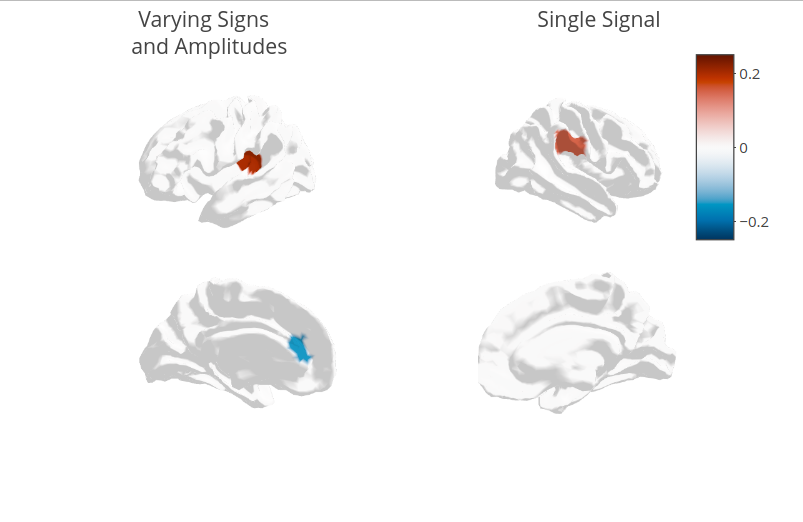}
\end{minipage}
\caption{Simulated signal plateau regions for both hemispheres. Colors indicate signal amplitudes $\tau_p$, with positive (red) and negative (blue) activations represented by the color scales. Left panel: scenarios \textit{varying signs} and \textit{varying amplitudes}; right panel: scenarios \textit{varying signs and amplitudes} and \textit{single signal}.} \label{fig:sim_signals}
\end{figure}

Figure~\ref{fig:sim_signals} illustrates the underlying signal configurations in the same order from left to right. Signal centers are placed such that the resulting regions are spatially separated on the cortical surface.

These varying configurations allow for the assessment of method performance under increasing structural complexity, with a single isolated signal as well as two signal regions on the cortical surface with differing characteristics. To additionally evaluate the influence of sample size, each setting is simulated for $N = 25$ and $N = 50$ subjects.

Figure~\ref{fig:sim_average} in the Appendix shows corresponding averaged measurements across subjects.

\subsection{Results}

For each signal configuration, 100 independent simulation runs are conducted for both $N = 25$ and $N = 50$ subjects. Cluster-based permutation tests are included as a benchmark for detection performance, using $1{,}024$ permutations and a two-sided significance level of $\alpha = 0.05$.

For the boosting approach, a step length of $\nu = 0.1$ is used. The number of boosting iterations is initialized at $m_{\text{stop}} = 3{,}000$ and optimized via 10-fold cross-validation, followed by deselection with $\kappa = 0.01$ to improve sparsity and interpretability.

\begin{figure}[!h]
\centering
\includegraphics[width=\textwidth]{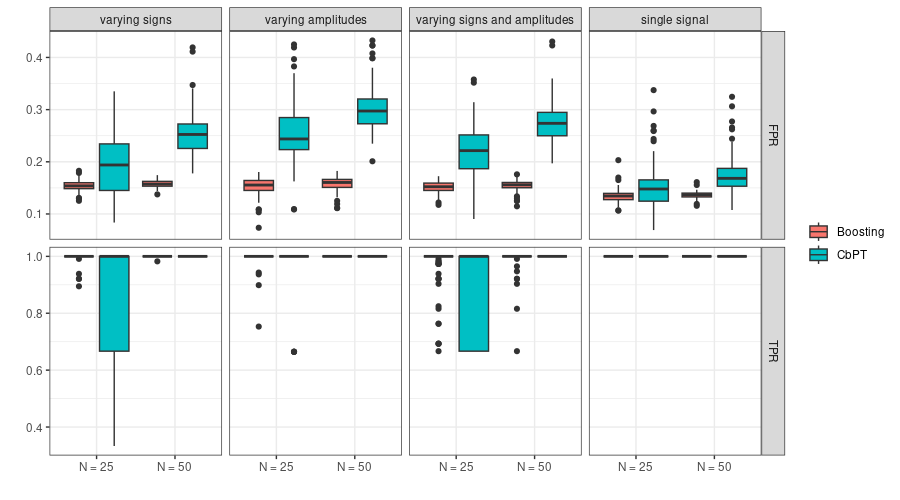}
\caption{Results for $FPR$ (first row) and $TPR$ (second row). The boosting approach is shown in red (\textit{Boosting}) and cluster-based permutation tests in blue (\textit{CbPT}). Columns correspond to the four simulation settings: varying signs, varying amplitudes, varying signs and amplitudes, and single signal.} \label{fig:sim_res_rates}
\end{figure}

Figure~\ref{fig:sim_res_rates} summarizes detection performance across all settings. With respect to $TPR$, cluster-based permutation tests (\textit{CbPT} in Figure legend) show reduced sensitivity in the \textit{varying signs} and \textit{varying signs and amplitudes} settings for $N = 25$, failing to detect parts of the true signal regions in several simulation runs. In contrast, the boosting approach achieves near-complete detection, with values below $1$ occurring only as outliers.

For the simplest configuration, \textit{single signal}, both methods reliably recover the full signal plateau for both sample sizes. Across all configurations, increasing the sample size to $N = 50$ improves detection performance for cluster-based permutation tests. However, this improvement coincides with a loss in spatial specificity (see below). 

In contrast, the boosting approach exhibits stable and consistent $TPR$ across all settings and sample sizes, where instances of missed vertices are rare, occur only in a small number of simulation runs, and are primarily observed in the smaller sample size scenarios.

For all simulation settings, the proposed boosting approach yields consistently lower $FPR$ values compared to cluster-based permutation tests. Even further, the boosting approach shows stable behavior, with $FPR$ values remaining within a range of approximately $10\%$--$20\%$ across all settings and sample sizes.

Conversely, cluster-based permutation tests exhibit increasing $FPR$ with larger sample sizes, indicating a tendency to expand detected regions as more data becomes available. This behavior aligns with known limitations of these methods, which do not provide valid spatial localization of effects and can lead to systematic overestimation of activation extent \citep{sassenhagen2019cptproblems, rousselet2025badCPT}. These differences in $FPR$ between the two approaches are particularly pronounced in the \textit{varying amplitudes} and \textit{varying signs and amplitudes} settings, where signal regions differ in amplitudes and in the latter one also in sign.

Table~\ref{tab:sim_errormetrics} reports estimation accuracy of the boosting approach. Across all settings and sample sizes, both $\text{mae}_\tau$ and $\text{rmse}_\tau$ indicate stable and accurate recovery of the underlying signal amplitudes. 
In the \textit{single signal} case both error metrics are smaller due to reduced structural complexity in the absence of overlapping halo regions.

In summary, the results highlight a clear trade-off between detection and spatial specificity. While cluster-based permutation tests achieve high detection rates for larger sample sizes, this comes at the cost of inflated spatial extent. For smaller sample sizes, they may fail to detect parts of the signal. In contrast, the proposed boosting approach maintains a favorable balance between detection power and spatial precision across all settings, while additionally providing accurate effect size estimates.

Section~\ref{fig:est_res} in the Appendix provides visual examples of the boosting-based estimates alongside significant clusters from the cluster-based permutation tests.

\begin{table}[h]
\centering
\begin{tabular}{ll|cc}
\toprule
Configuration & N & $\text{mae}_\tau$ ($\pm$ sd) & $\text{rmse}_\tau$ ($\pm$ sd) \\
\midrule
varying signs & 25 & 0.0110 ($\pm$ 0.0013) & 0.0371 ($\pm$ 0.0040)\\
& 50 & 0.0111 ($\pm$ 0.0009) & 0.0370 ($\pm$ 0.0027)\\
varying amplitudes & 25 & 0.0112 ($\pm$ 0.0010) & 0.0379 ($\pm$ 0.0030)\\
& 50 & 0.0113 ($\pm$ 0.0008) & 0.0380 ($\pm$ 0.0021)\\
\midrule
varying signs & 25 & 0.0133 ($\pm$ 0.0013) & 0.0450 ($\pm$ 0.0043)\\
and amplitudes & 50 & 0.0132 ($\pm$ 0.0009) & 0.0446 ($\pm$ 0.0028)\\
single signal & 25 & 0.0093 ($\pm$ 0.0015) & 0.0320 ($\pm$ 0.0049)\\
& 50 & 0.0095 ($\pm$ 0.0011) & 0.0326 ($\pm$ 0.0036) \\
\bottomrule
\end{tabular}
\caption{Estimation accuracy of the boosting approach across all simulation configurations.}
\label{tab:sim_errormetrics}
\end{table}

%% file: 04data.tex
\section{Experiment Application}
\label{sec:data}

To illustrate the proposed methodology on experimental data, an analysis of a dataset consisting of $N = 23$ patients diagnosed with tinnitus follows. The participants performed alternating relaxing and straining jaw exercises for a duration of 90 seconds. Directly after task performance, oscillatory brain activity was recorded for three minutes, while participants fixated a black cross in the middle of a screen. This analysis focuses on time-averaged activity in the gamma frequency range (32\,--\,89 Hz). The study was approved by the ethics committee of the University Hospital Erlangen. For further details on the data and their acquisition, see \cite{sabados2025experiment}.

Source reconstruction was performed to project the sensor-level measurements onto the cortical surface. The resulting data are represented on the white-matter surface of a template brain using the \textit{fsaverage} surface with oct6 spacing. This yields $n_v = 4{,}098$ cortical vertices per hemisphere, such that the whole setting is consistent with the simulation setup. The spatial analyses are conducted separately for each hemisphere.

For each subject, centered differences in oscillatory activity between the strained and relaxed conditions are computed at each vertex, yielding a subject-specific spatial field of condition effects. These centered differences serve as the response variable in the subsequent analyses.

\begin{figure}[htbp]
\centering
\begin{minipage}{0.55\textwidth}
  \centering
  \includegraphics[trim={4cm 4cm 1cm 0},clip,width=0.8\textwidth]{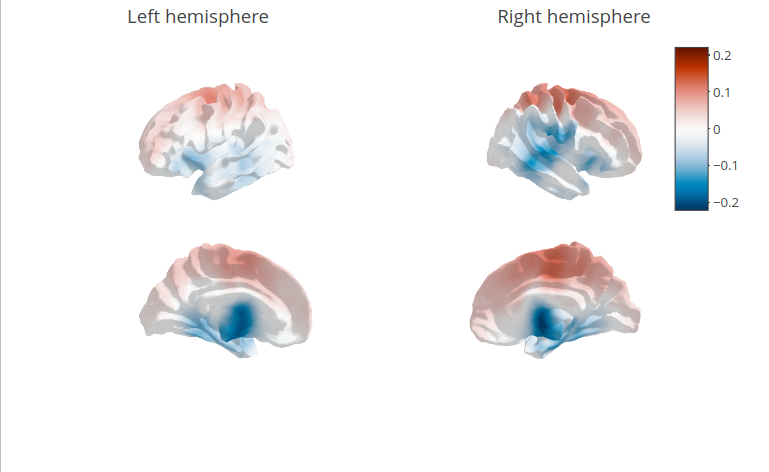}
\end{minipage}\hfill
\begin{minipage}{0.43\textwidth}
  \centering
  \includegraphics[trim={3cm 3cm 1cm 0},clip,width=0.8\textwidth]{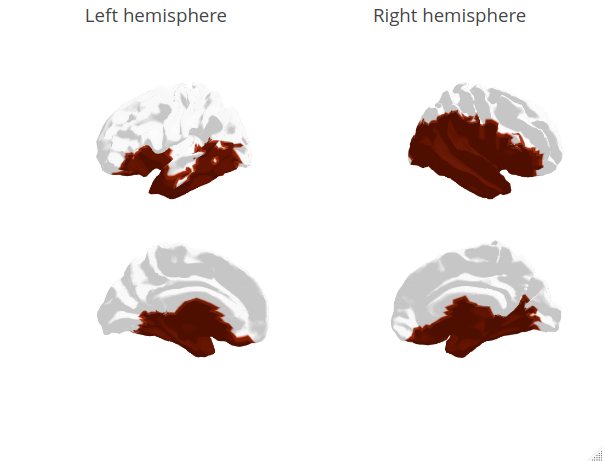}
\end{minipage}
\caption{Left panel: Observed centered differences of gamma-band activity between conditions, averaged across subjects; right panel: Results from cluster permutation tests on experimental data.} \label{fig:data_current}
\end{figure}

Figure~\ref{fig:data_current} (left panel) shows the average condition difference across all subjects. While spatial patterns of increased and decreased gamma activity are visible, substantial variability remains at the individual level (see Appendix~\ref{sec:data_ind_brains}).

Again, cluster-based permutation tests are applied as a benchmark analysis. Inference is based on $2,048$ permutations with a two-sided significance level of $\alpha = 0.05$. The resulting significant clusters are shown in the right panel of Figure~\ref{fig:data_current}.

In total, $13$ clusters are detected across both hemispheres, several of which include only a few vertices. However, two large clusters dominate the results: one cluster comprising $1,044$ vertices in the left hemisphere and one cluster comprising $1,906$ vertices in the right hemisphere. After permutation-based inference, these two large clusters are found to be significant. In the left hemisphere, this cluster is primarily located in inferior cortical regions, whereas the significant cluster in the right hemisphere covers a substantial portion of the cortical surface, indicating the presence of spatially extended effects but providing limited spatial specificity.

The proposed gradient boosting approach based on WRBF is tuned using leave-two-out cross-validation, resulting in an optimal stopping iteration of $m^* = 447$. Again, the described deselection approach is applied according to the simulation setup.

\begin{figure}[htbp]
\centering

\begin{minipage}{0.48\textwidth}
  \centering
  \includegraphics[trim={2cm 3cm 1.5cm 0.3cm},clip,width=\textwidth]{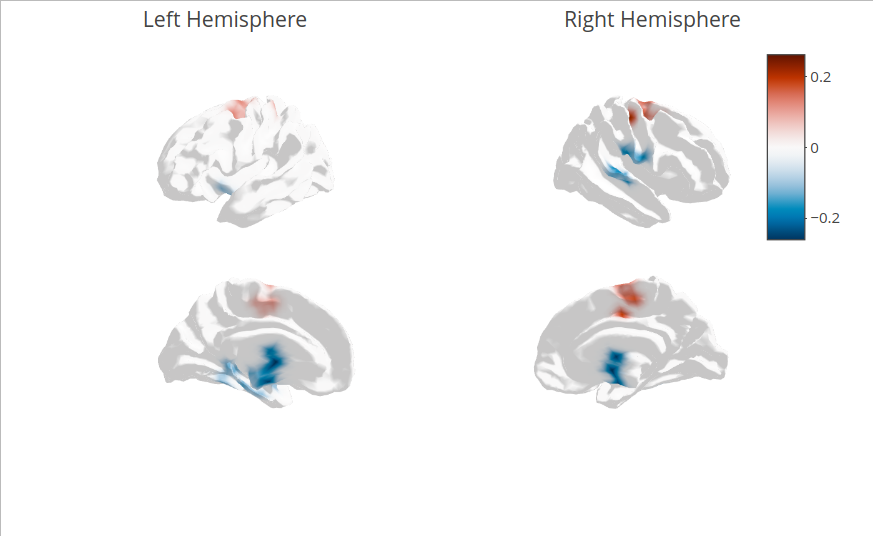}
\end{minipage}\hfill
\begin{minipage}{0.48\textwidth}
  \centering
  \includegraphics[trim={2cm 3cm 1.5cm 0.3cm},clip,width=\textwidth]{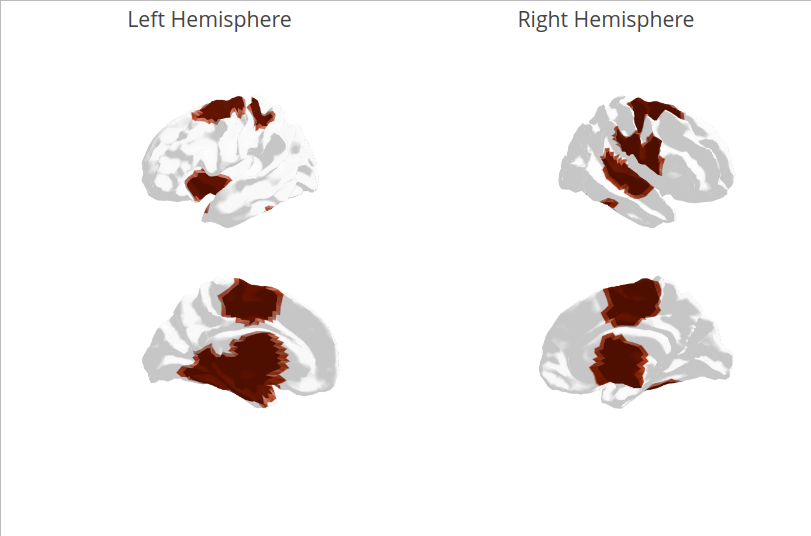}
\end{minipage}
\caption{Empirical results from gradient boosted WRBF. Left panel: effect estimation; right panel: spatial regions covered by WRBF.} \label{fig:data_WendlandBoost}
\end{figure}

In the left hemisphere, $1{,}059$ vertices (443 vertices with an estimated positive effect, $\widehat{f(\boldsymbol{s}_v)} > 0$; 616 vertices with an estimated negative effect, $\widehat{f(\boldsymbol{s}_v)} < 0$) are assigned non-zero effect estimates, while in the right hemisphere $933$ vertices (412 vertices with $\widehat{f(\boldsymbol{s}_v)} > 0$; 512 vertices with $\widehat{f(\boldsymbol{s}_v)} < 0$) are covered by selected basis functions. The right panel of Figure~\ref{fig:data_WendlandBoost} displays these selected regions, i.e., vertices covered by at least one selected basis function. This enables a direct comparison to the cluster permutation results in terms of spatial extent. The corresponding estimated spatial effects are shown in the left panel.

Compared to the cluster permutation results, the selected regions are generally more localized. Although the selected region in the left hemisphere remains relatively large, the estimated effects provide a more differentiated picture, revealing negative effects in inferior regions and positive effects in superior areas. This pattern is consistent with the average condition differences shown in Figure~\ref{fig:data_current}.

In the right hemisphere, several spatially distinct regions with varying effect magnitudes are identified. In particular, superior cortical areas exhibit smaller effect sizes and opposing signs relative to inferior regions. These localized patterns are captured by the boosting approach through spatially resolved effect estimates, allowing regions with differing magnitudes and directions to be distinguished.

Such heterogeneous signal configurations are not reflected in the cluster permutation results, where large clusters aggregate signals across space. In line with this, no significant cluster is detected in the superior cortical regions. Regions with smaller effect sizes or opposing signs may not contribute sufficiently to cluster-level statistics and therefore remain undetected. This behavior is consistent with the simulation scenarios, where heterogeneous signal configurations led to reduced detection sensitivity for cluster-based methods.

This spatially resolved estimation allows for a more nuanced interpretation of the data. Gamma power seems to be mainly reduced in right cortical regions encompassing the right auditory and the right somatosensory cortex (jaw area). This pattern may reflect a reduction in tinnitus-associated auditory activity \citep{weisz2007neural} together with reduced activity in regions related to jaw sensation, potentially pointing to a continuing 'relaxed feeling in the jaw' \citep{yordanova2025eeg} following the relaxing jaw exercises. These is consistent with the hypotheses formulated in \cite{sabados2025experiment}. 

The revealed left and right medial sources are localized in deeper brain regions, where the exact localization of neuronal activity from scalp-based measurements is limited, so that we would refrain from a further interpretation. 

In addition, the gradient boosted WRBF approach reveals increased gamma activity in the left and right central brain regions. These regions overlap with areas showing increased alpha activity in \cite{sabados2025experiment}, which may indicate a more complex interaction between oscillatory power in different frequency bands within the sensori-motor cortex.

Overall, the application reflects the findings from the simulation study. Cluster permutation tests tend to favor spatially extended clusters, potentially combining regions with heterogeneous signal characteristics. In contrast, the boosting approach emphasizes localized effects and allows for varying signal magnitudes and directions across space.

%% file: 05conclusion.tex
\section{Conclusion}
\label{sec:conclusion}

This work introduces a unified statistical framework for the analysis of spatial brain activity on the cortical surface. The proposed approach reformulates the analysis of experimental MEG data within a functional regression setting, enabling direct estimation as well as selection of spatial activation effects rather than relying solely on hypothesis testing.

By doing so, a model-based representation of source-reconstructed MEG data which is compatible with regression methodology is provided . Spatial effects are modeled directly on the cortical surface using Wendland radial basis functions, allowing for flexible yet localized representations that respect the underlying mesh geometry. For data-driven selection and estimation, model-based gradient boosting is employed, combining statistical learning with interpretability. Finally, a comprehensive simulation framework is introduced that generates realistic cortical activation patterns, allowing systematic evaluation of methodological performance under controlled conditions.

The empirical results and simulations highlight that the proposed framework yields spatially resolved effect estimates and reveals heterogeneous activation patterns. In particular, signal regions with smaller effect sizes or opposing signs can be difficult to detect using cluster-based methods, as demonstrated in the simulations, but are directly captured within the regression-based approach. At the same time, the simulation results show that this improved detection is achieved without inflating the spatial extent of identified regions, resulting in a consistent balance between accurate localization and reliable detection across different signal configurations and sample sizes.

A comparison against a Bayesian spatial GLM adapted from an existing fMRI-based framework (Appendix~\ref{app:bayesglm}) further corroborates these findings, showing consistently higher and more variable estimation errors across all simulation settings. Its post-hoc excursion-based selection additionally depends on a manually chosen effect-size threshold that induces a pronounced trade-off between detection and false positive rates, in contrast to the primarily data-driven, early-stopping-based regularization employed here. Being developed for repeated fMRI measurements, the method relies on comparably larger sample sizes for reliable detection performance, in contrast to the more robust boosting approach.

Beyond detection and effect estimation, the proposed framework enables a range of methodological extensions. As a regression-based model, it allows for the straightforward inclusion of additional covariates, interactions, or subject-specific effects, and can be extended to alternative model classes such as mixed-effects models \citep{breslow1993glmms} or distributional regression frameworks (e.g., GAMLSS, see \citeauthor{rigby2005gamlss}, \citeyear{rigby2005gamlss}). Both model classes are already discussed in the context of gradient boosting \citep{knieper2025gbforgamm, mayr2010gamlss}. Moreover, the approach does not require prior assumptions about the spatial location of effects, relying instead on assumptions about the outcome distribution (oscillatory power measurements), which can be assessed in preliminary analyses.

An important aspect of the proposed method is the use of early stopping in gradient boosting as a data-driven regularization mechanism. Model complexity is controlled via predictive performance, which mitigates overfitting and reduces the need for extensive multiple testing corrections. While the manually chosen threshold parameter $\kappa$ is introduced in the deselection step, its influence is limited and primarily serves to improve interpretability by removing negligible contributions.

Despite these advantages, several limitations remain and provide directions for future research. The present analysis focuses on a single frequency band, whereas extensions to multivariate outcomes or joint modeling across frequencies may yield additional insights. Furthermore, the current framework operates on time-averaged data. Extending the model to incorporate the temporal dimension would allow investigation of not only where but also when effects occur, addressing questions raised in previous work on cluster-based inference \citep{sassenhagen2019cptproblems}. Such an extension would also enable a genuinely appropriate comparison against the Bayesian alternative of \citet{Mejia02042020}, whose original pipeline presupposes exactly this kind of temporal structure. Unlike the corresponding post-hoc, excursion-based selection, gradient boosting would extend its inherently data-driven selection naturally into the temporal dimension.

In summary, the proposed approach shifts the analysis of cortical surface data from purely inferential procedures toward regression-based modeling, enabling interpretable, spatially resolved effect estimation. It provides a flexible and extensible framework for analyzing cortical surface data, with the potential to yield more detailed insights into the spatial structure of brain activity.

\clearpage

%% file: 06appendix.tex
\section{Appendix}

\subsection{Geodesic distance on the cortical surface}
\label{sec:geodesic}

Spatial relationships between cortical vertices are quantified using geodesic distances defined along the cortical surface mesh throughout this work. In contrast to Euclidean distances in three-dimensional space, geodesic distances respect the intrinsic geometry of the cortical surface and therefore provide a more appropriate measure of spatial proximity for surface-based brain analyses.

The cortical surface is represented as a triangular mesh consisting of vertices connected by edges.
The geodesic distance between two vertices $v$ and $v'$ is defined as the length of the shortest path along the mesh that connects them. That is, among all possible paths that move from vertex to vertex along the surface, the geodesic distance is the minimal total length obtained by summing the distances of consecutive edges along the path.

In practice, these distances are computed using shortest-path algorithms on the mesh, where edge lengths are given by Euclidean distances between neighboring vertices. This approach provides an efficient and accurate approximation of intrinsic surface distances \citep{kimmel1998computinggeodesic,surazhsky2005geodesicmesh}.

\subsection{Spectral generation of spatial Gaussian noise}
\label{app:GaussianNoise}

To generate spatially correlated Gaussian noise fields on the cortical surface, we simulate realizations from a multivariate normal distribution with covariance matrix $\sigma^2\mathbf{K}$, where $\mathbf{K} \in \mathbb{R}^{n_v \times n_v}$ is defined by the spatial kernel

$$
K_{vv'} =
\exp\!\left(
-\frac{d_G(v,v')^2}{2\ell^2}
\right),
$$

with $d(v,v')$ denoting the geodesic distance between vertices $v$ and $v'$ along the cortical surface and $\ell$ controlling the spatial correlation range.

Direct sampling from $\mathcal{N}(\mathbf{0},\sigma^2\mathbf{K})$ can be performed using a Cholesky factorization of the covariance matrix. However, repeatedly computing such matrix factorizations becomes computationally expensive for large surface meshes.

Instead, we employ a spectral representation of the covariance matrix. Let

$$
\mathbf{K} =
\mathbf{U}
\boldsymbol{\Lambda}
\mathbf{U}^{\top}
$$

denote the eigen-decomposition of $\mathbf{K}$, where $\mathbf{U}$ contains the eigenvectors and $\boldsymbol{\Lambda}$ is a diagonal matrix containing the
corresponding eigenvalues. A spatial Gaussian noise vector can then be generated as

$$
\boldsymbol{\varepsilon}
=
\sigma
\mathbf{U}
\boldsymbol{\Lambda}^{1/2}
\mathbf{z},
\quad
\mathbf{z} \sim
\mathcal{N}(\mathbf{0}, \mathbf{I}_{n_v}),
$$

which generates $\boldsymbol{\varepsilon} \sim \mathcal{N}(\mathbf{0},\sigma^2\mathbf{K})$.

In practice, the eigen-decomposition of $\mathbf{K}$ is computed once for each hemisphere prior to the simulation study. Subsequent noise realizations can then be generated efficiently using only matrix--vector multiplications, which substantially reduces computational cost when simulating multiple subjects. To ensure numerical stability of the decomposition, a small diagonal regularization term is added to $\mathbf{K}$ prior to eigen-decomposition.

\clearpage

\subsection{Simulation: Individual Brain Effects}
\label{sec:App_sim_measurements}

\begin{figure}[h!]
\centering

\begin{minipage}{0.48\textwidth}
  \centering
  \includegraphics[trim={4cm 3cm 0.9cm 0.3cm},clip,width=\linewidth]{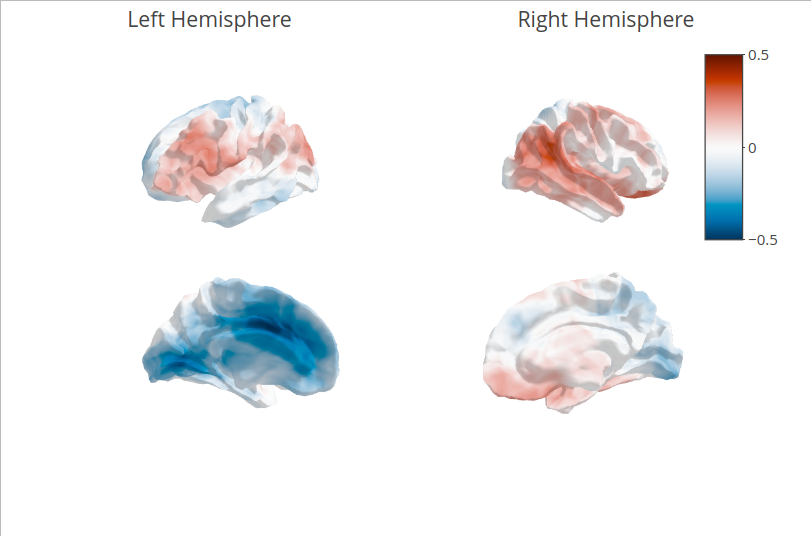}
\end{minipage}\hfill
\begin{minipage}{0.48\textwidth}
  \centering
  \includegraphics[trim={4cm 3cm 0.9cm 0.3cm},clip,width=\linewidth]{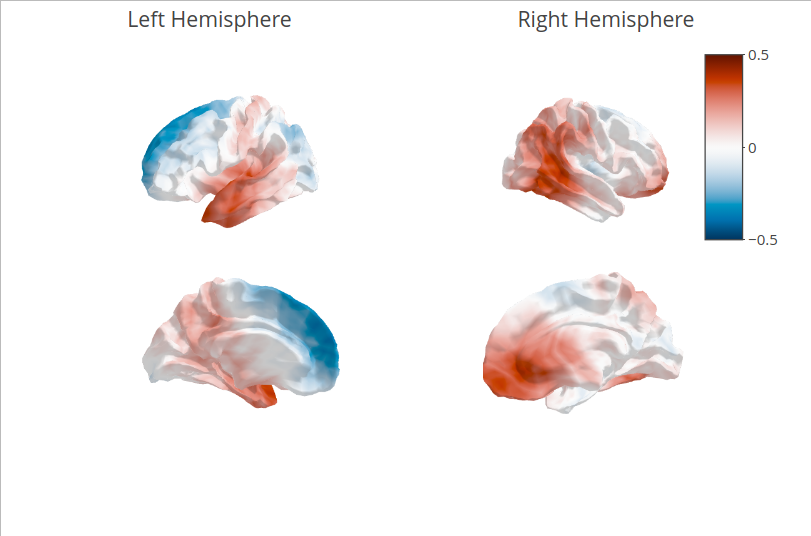}
\end{minipage}

\begin{minipage}{0.48\textwidth}
  \centering
  \includegraphics[trim={4cm 3cm 0.9cm 0.3cm},clip,width=\linewidth]{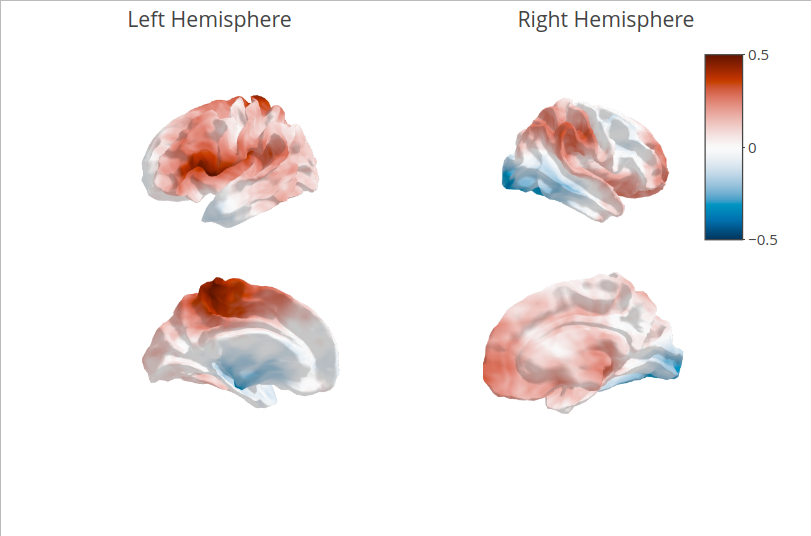}
\end{minipage}\hfill
\begin{minipage}{0.48\textwidth}
  \centering
  \includegraphics[trim={4cm 3cm 0.9cm 0.3cm},clip,width=\linewidth]{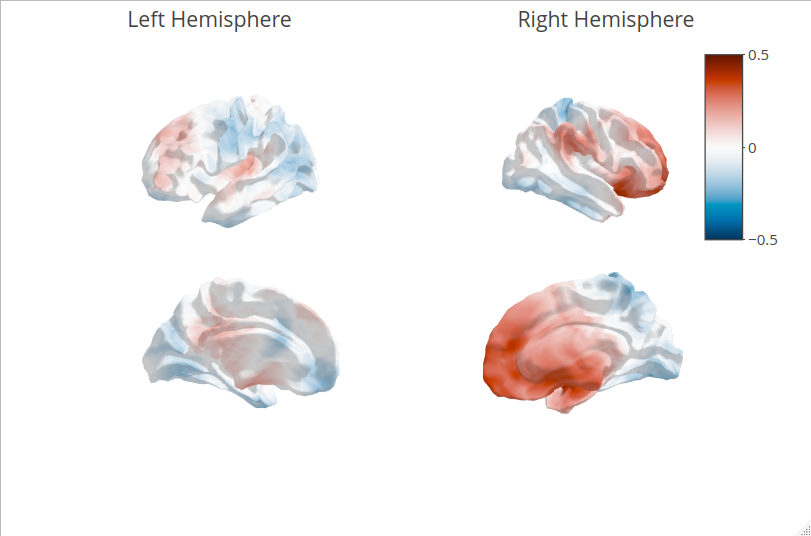}
\end{minipage}

\begin{minipage}{0.48\textwidth}
  \centering
  \includegraphics[trim={4cm 3cm 0.9cm 0.3cm},clip,width=\linewidth]{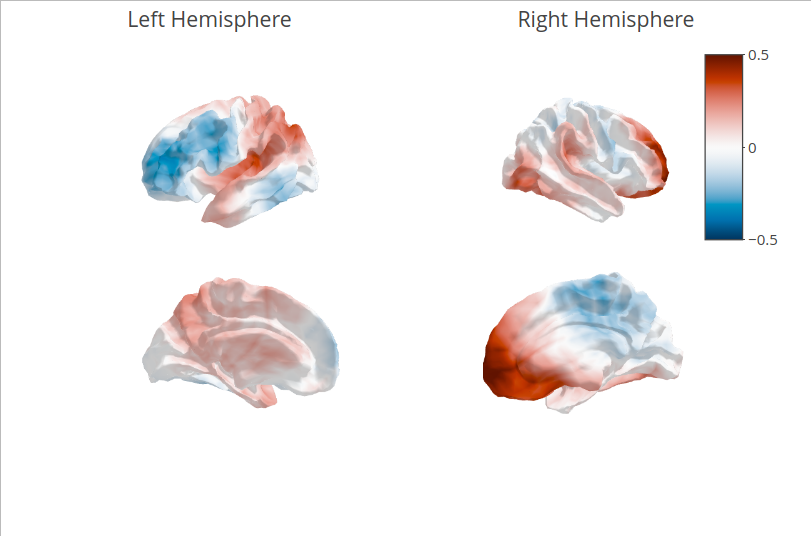}
\end{minipage}\hfill
\begin{minipage}{0.48\textwidth}
  \centering
  \includegraphics[trim={4cm 3cm 0.9cm 0.3cm},clip,width=\linewidth]{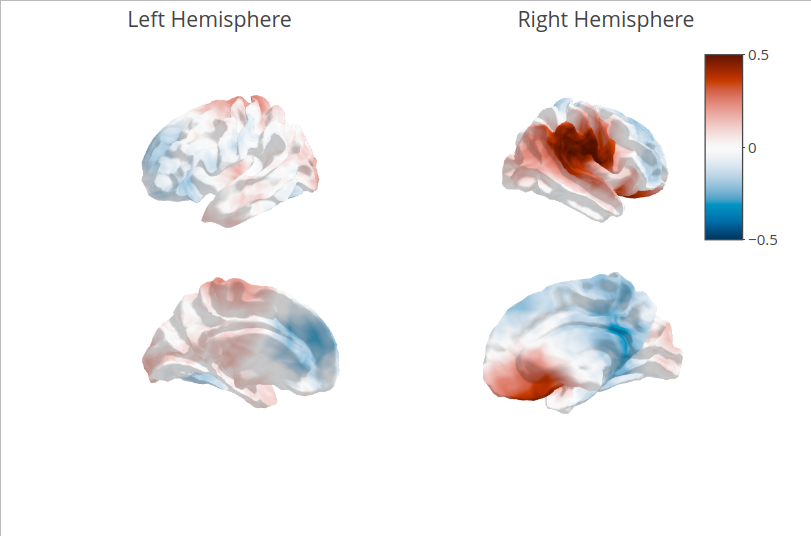}
\end{minipage}

\begin{minipage}{0.48\textwidth}
  \centering
  \includegraphics[trim={4cm 3cm 0.9cm 0.3cm},clip,width=\linewidth]{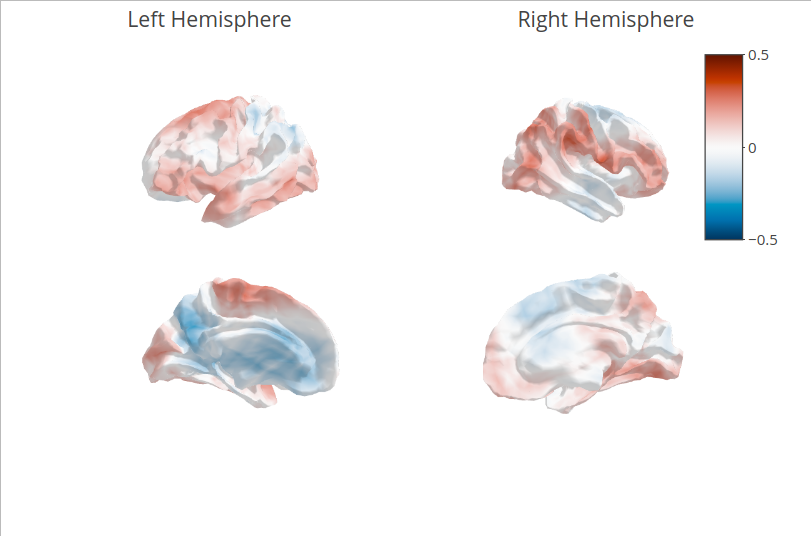}
\end{minipage}\hfill
\begin{minipage}{0.48\textwidth}
  \centering
  \includegraphics[trim={4cm 3cm 0.9cm 0.3cm},clip,width=\linewidth]{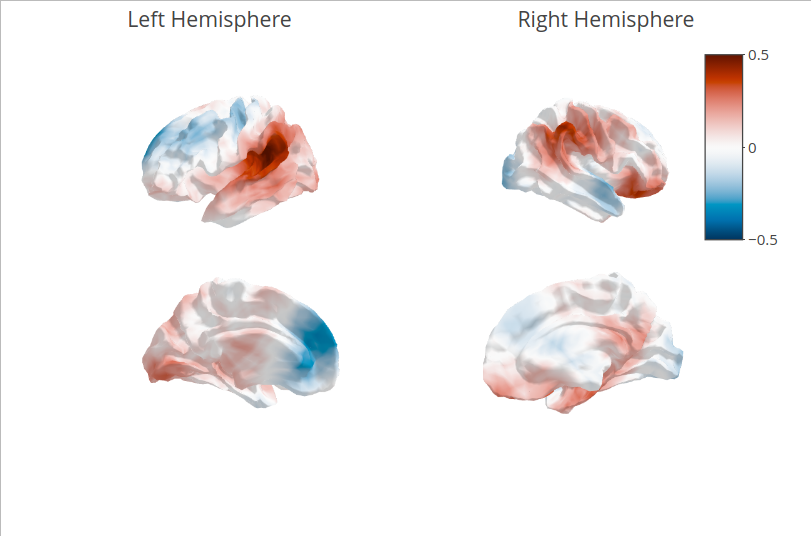}
\end{minipage}
\caption{Examples of subject-level effects from the simulation setting varying signs and varying amplitudes.} 
\end{figure}

\clearpage

\begin{figure}[htbp]
\centering
\begin{minipage}{0.48\textwidth}
  \centering
  \includegraphics[trim={3cm 3cm 1cm 0.3cm},clip,width=\linewidth]{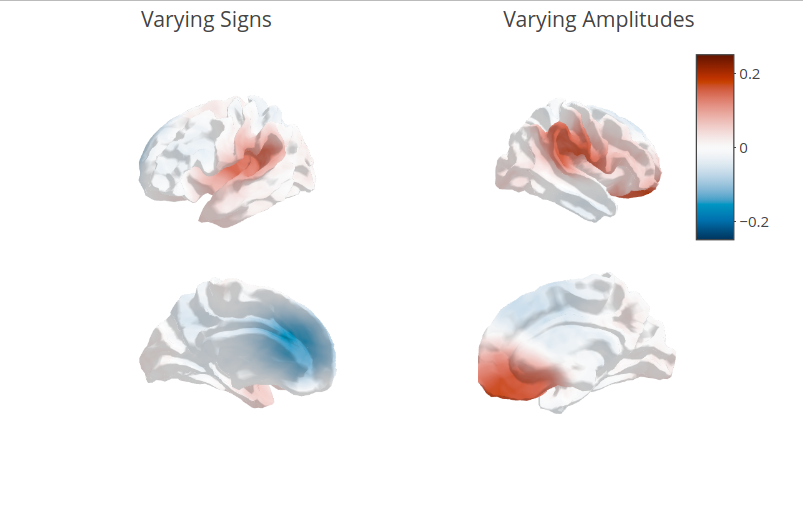}
\end{minipage}\hfill
\begin{minipage}{0.48\textwidth}
  \centering
  \includegraphics[trim={3cm 3cm 1cm 0.3cm},clip,width=\linewidth]{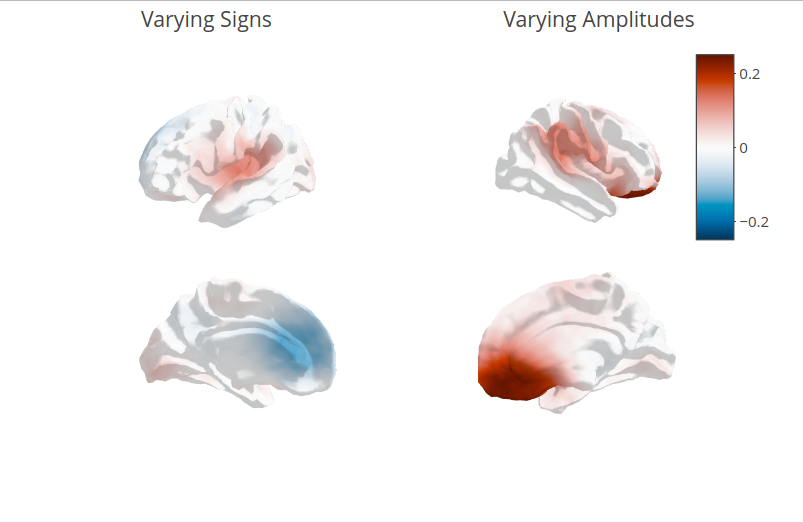}
\end{minipage}

\vspace{0.5em}

\begin{minipage}{0.48\textwidth}
  \centering
  \includegraphics[trim={3cm 3cm 1cm 0.3cm},clip,width=\linewidth]{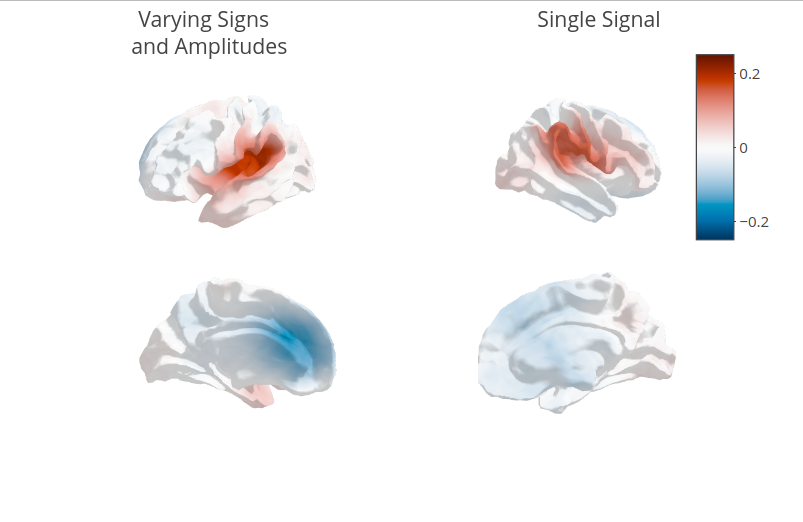}
\end{minipage}\hfill
\begin{minipage}{0.48\textwidth}
  \centering
  \includegraphics[trim={3cm 3cm 1cm 0.3cm},clip,width=\linewidth]{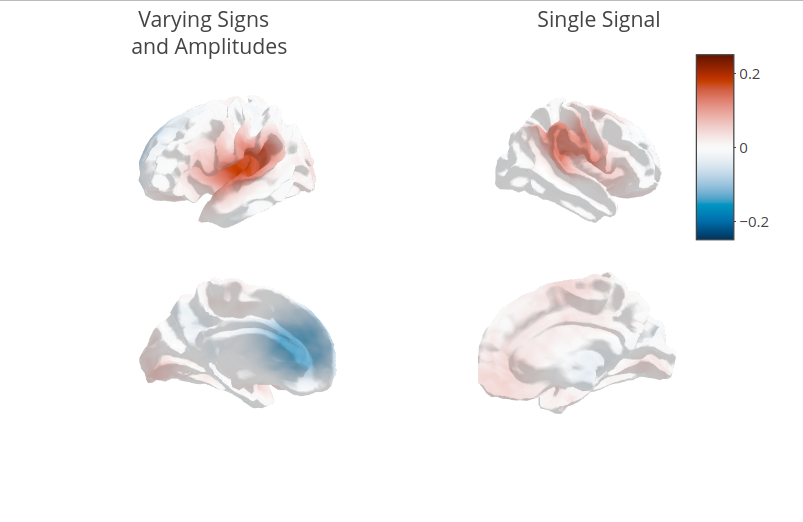}
\end{minipage}
\caption{Examples of simulated measurements averaged across $N = 25$ subjects. Each column represents another simulated experiment. First row: Scenarios \textit{varying signs} and \textit{varying amplitudes}; second row: scenario \textit{varying signs and amplitudes} and \textit{single signal}.} \label{fig:sim_average}
\end{figure}

\begin{figure}[htbp]
\centering
\begin{minipage}{0.49\textwidth}
  \centering
  \includegraphics[trim={3cm 3cm 1cm 0.3cm},clip,width=\textwidth]{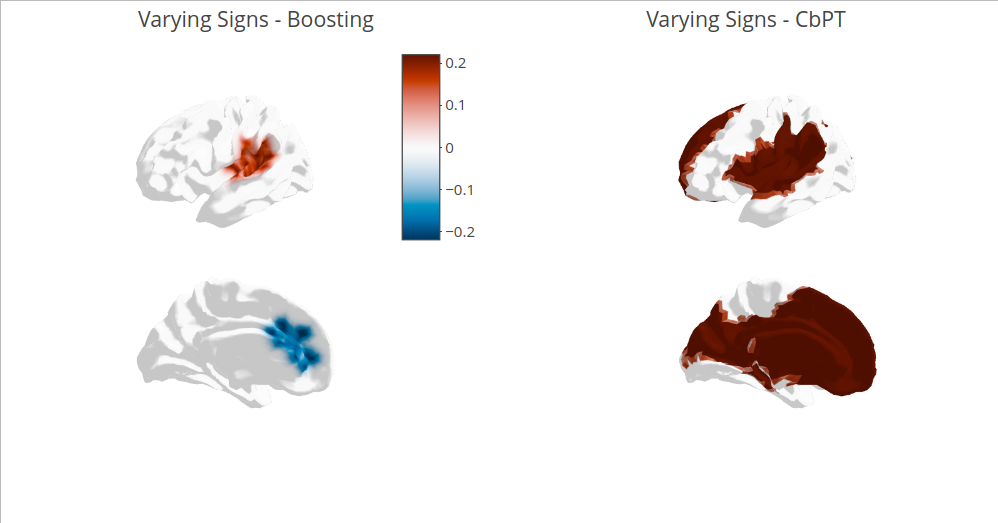}
\end{minipage}\hfill
\begin{minipage}{0.48\textwidth}
  \centering
  \includegraphics[trim={3cm 3cm 1cm 0.3cm},clip,width=\textwidth]{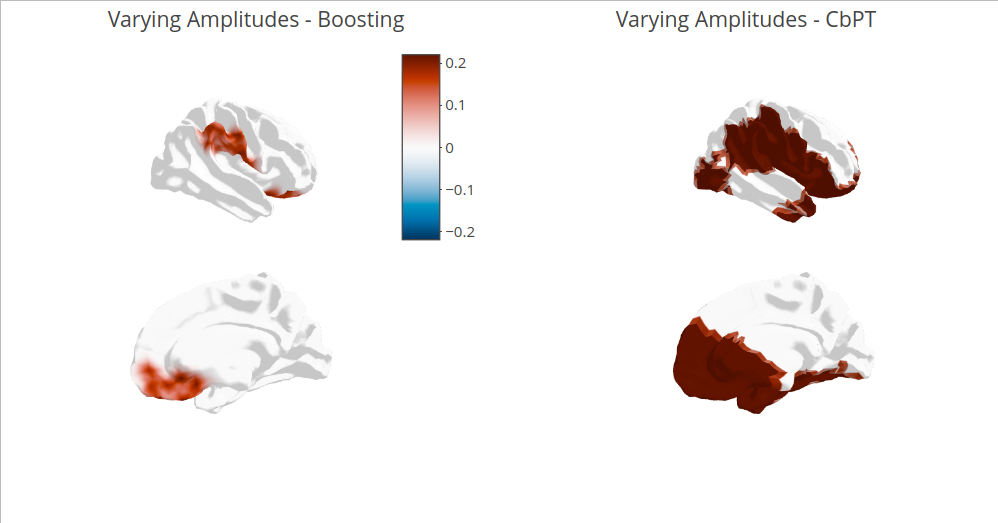}
\end{minipage}

\vspace{0.2em}

\begin{minipage}{0.49\textwidth}
  \centering
  \includegraphics[trim={3cm 3cm 1cm 0.3cm},clip,width=\textwidth]{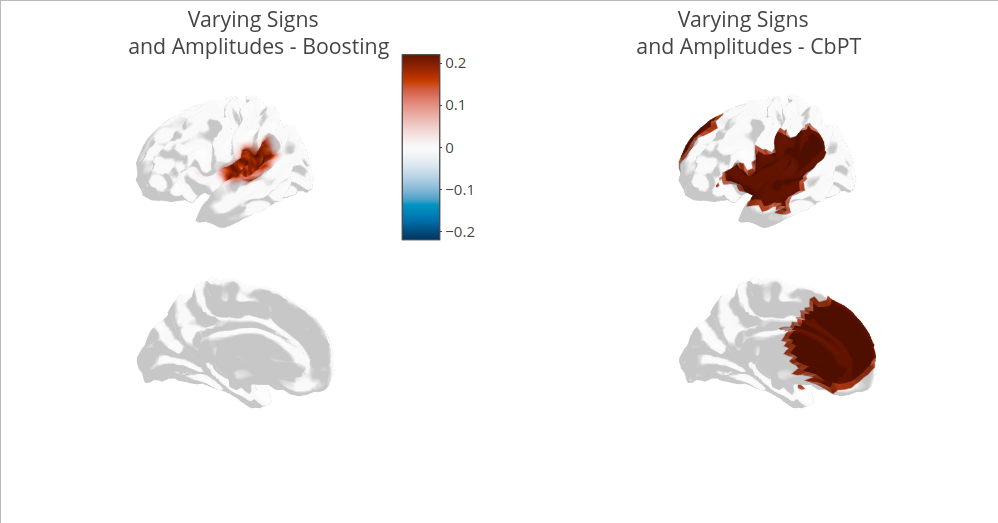}
\end{minipage}\hfill
\begin{minipage}{0.48\textwidth}
  \centering
  \includegraphics[trim={3cm 3cm 1cm 0.3cm},clip,width=\textwidth]{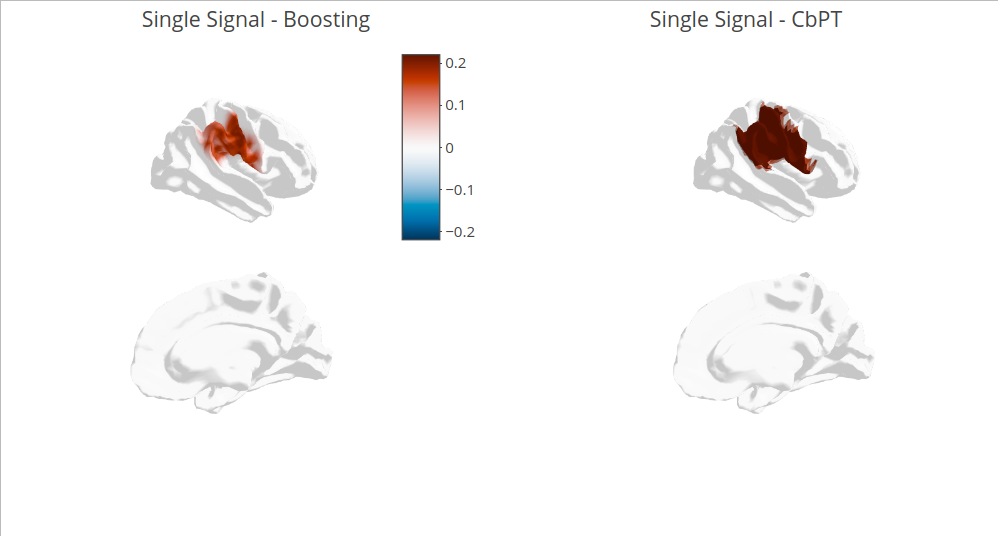}
\end{minipage}
\caption{One example result from each simulation setting highlighting far more detailed results of the proposed boosting approach. \\
\textit{Varying signs}: high $FPR$ for cluster-based permutation test (0.419) vs. only 0.157 for boosting).\\
\textit{Varying amplitudes}: high $FPR$ for cluster-based permutation test (0.424) vs. only 0.162 for boosting).\\
\textit{Varying signs and amplitudes}: Rare example of boosting missing one signal (TPR: 0.67) and cluster-based permutation testing detects signal relatively well.\\
\textit{Single signal:} Identical results in terms of $TPR$ (both 1) and $FPR$ (both 0.132) .}\label{fig:est_res}
\end{figure}

\clearpage

\subsection{Experiment: Individual Brain Differences}
\label{sec:data_ind_brains}

\begin{figure}[htbp]
\centering

\begin{minipage}{0.48\textwidth}
  \centering
  \includegraphics[trim={5cm 3cm 1cm 0},clip,width=\linewidth]{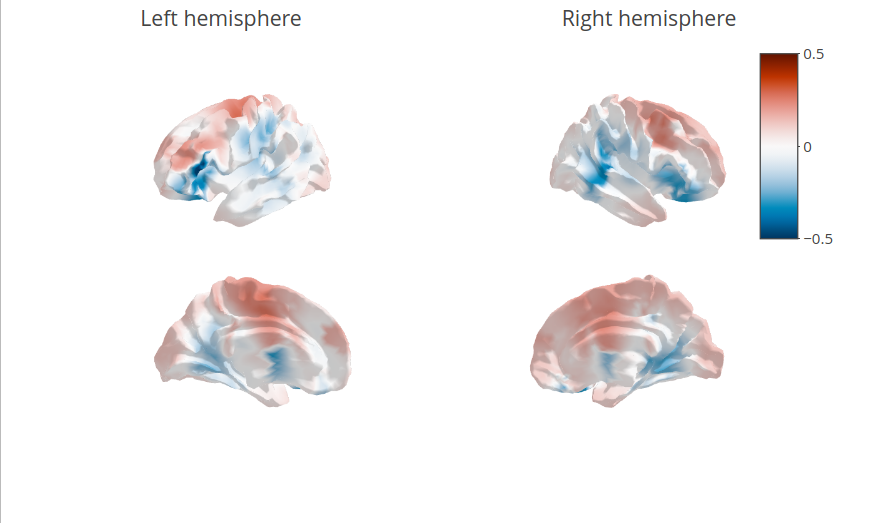}
\end{minipage}\hfill
\begin{minipage}{0.48\textwidth}
  \centering
  \includegraphics[trim={5cm 3cm 1cm 0},clip,width=\linewidth]{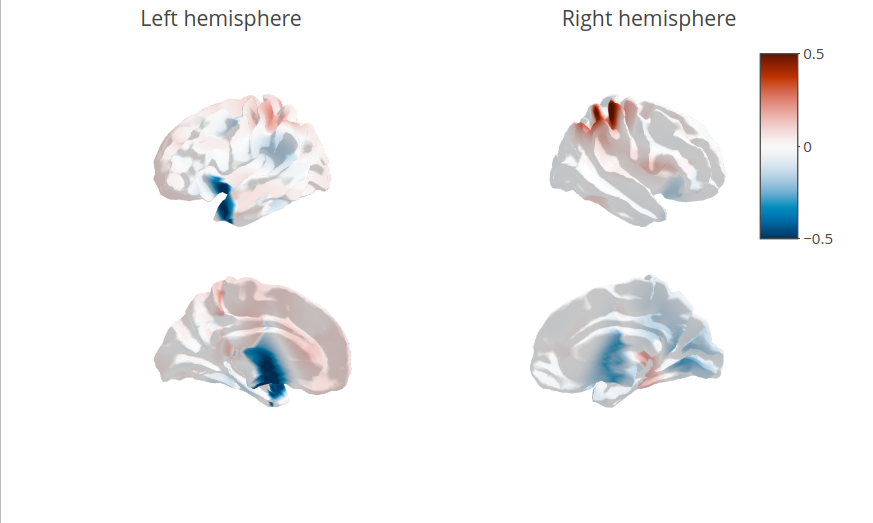}
\end{minipage}

\vspace{0.5em}

\begin{minipage}{0.48\textwidth}
  \centering
  \includegraphics[trim={5cm 3cm 1cm 0},clip,width=\linewidth]{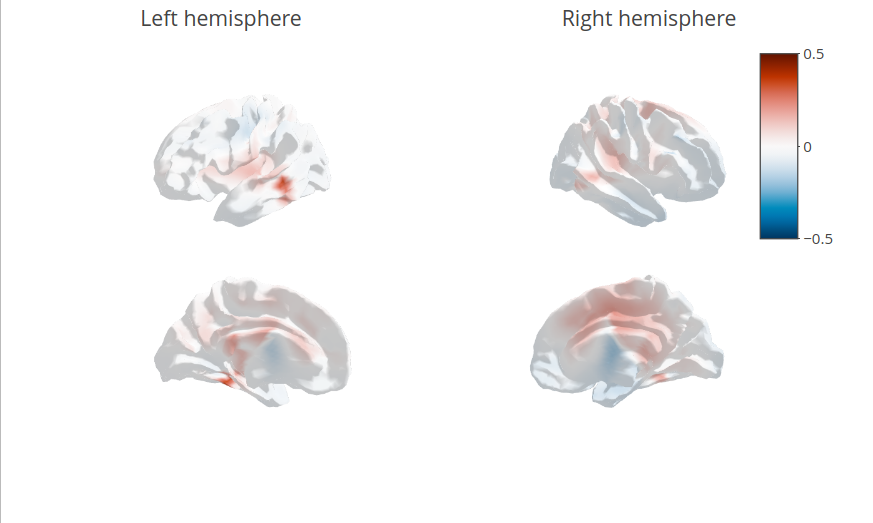}
\end{minipage}\hfill
\begin{minipage}{0.48\textwidth}
  \centering
  \includegraphics[trim={5cm 3cm 1cm 0},clip,width=\linewidth]{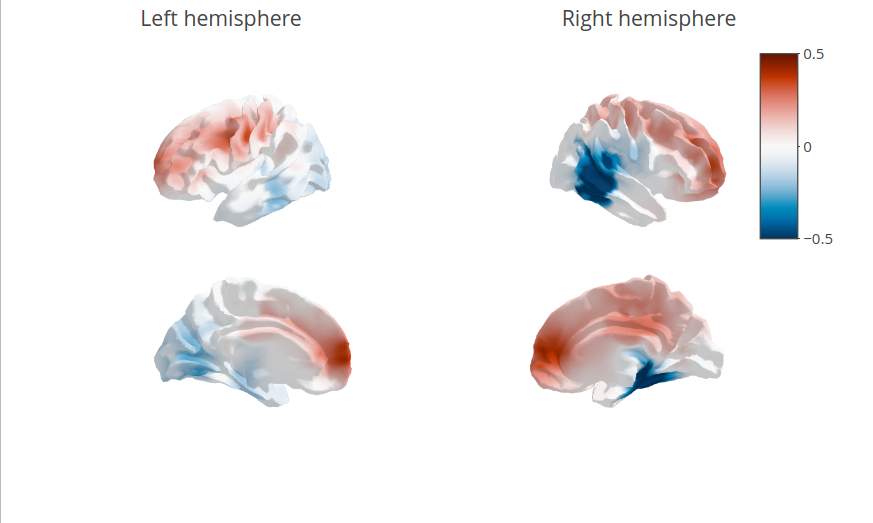}
\end{minipage}

\vspace{0.5em}

\begin{minipage}{0.48\textwidth}
  \centering
  \includegraphics[trim={5cm 3cm 1cm 0},clip,width=\linewidth]{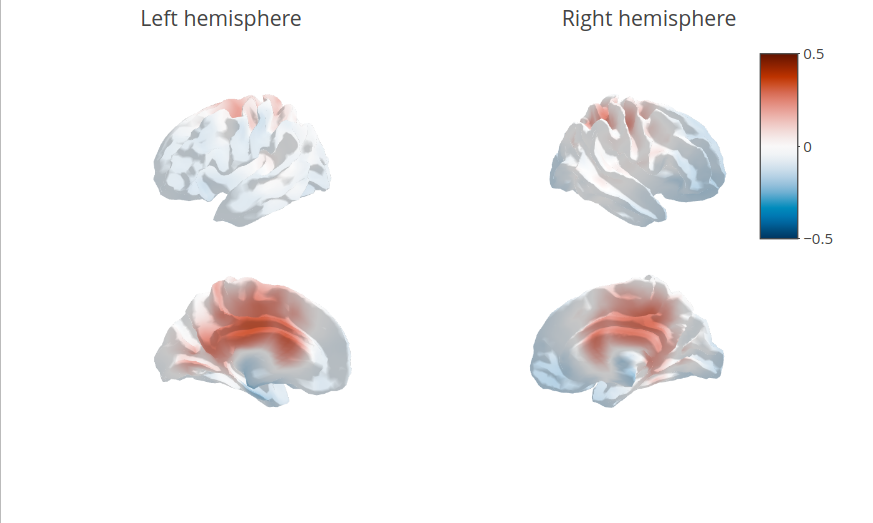}
\end{minipage}\hfill
\begin{minipage}{0.48\textwidth}
  \centering
  \includegraphics[trim={5cm 3cm 1cm 0},clip,width=\linewidth]{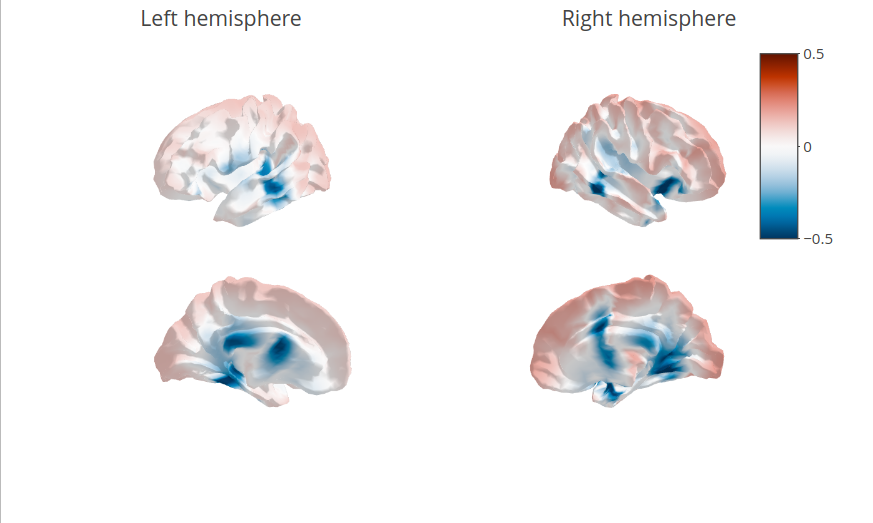}
\end{minipage}

\vspace{0.5em}

\begin{minipage}{0.48\textwidth}
  \centering
  \includegraphics[trim={5cm 3cm 1cm 0},clip,width=\linewidth]{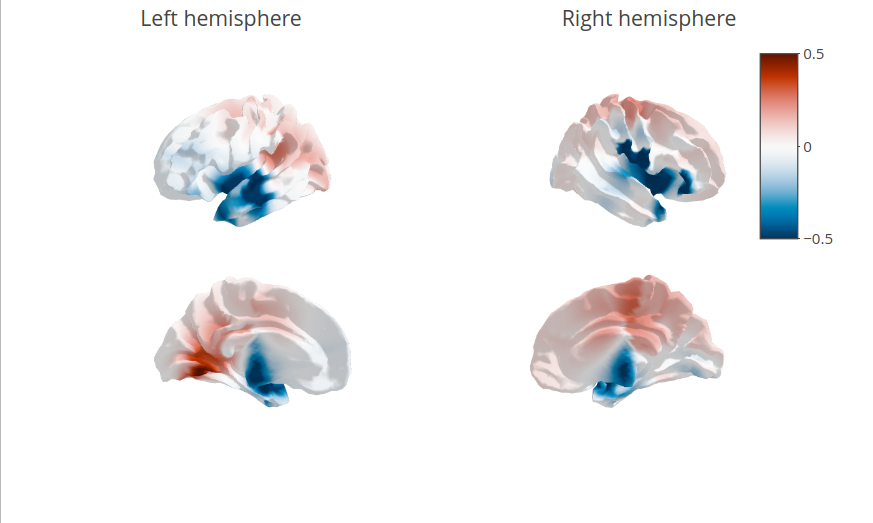}
\end{minipage}\hfill
\begin{minipage}{0.48\textwidth}
  \centering
  \includegraphics[trim={5cm 3cm 1cm 0},clip,width=\linewidth]{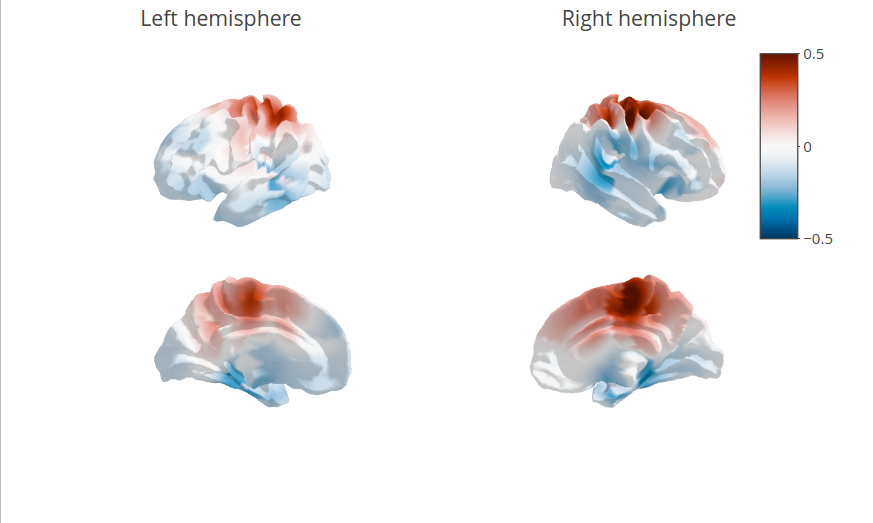}
\end{minipage}
\caption{Examples of subject-level centered difference measurements from real world experiment.} 
\end{figure}

\newpage
\input{BayesGLM}

%% file: BayesGLM.tex
\section{Bayes GLM}
\label{app:bayesglm}

\subsection{Methodological Adaptation}

The Bayes GLM approach of \citet{Mejia02042020} models repeated fMRI measurements on the cortical surface. To improve on the classical GLM, a spatial process prior is placed on the activation amplitude at each location. This prior builds on the stochastic partial differential equation (SPDE) approach of \citet{LINDGREN2022100599}: a continuous Mat\'ern Gaussian process is represented by a Gaussian Markov Random Field (GMRF) with a sparse precision matrix, defined on a triangular mesh of the cortical surface. Bayesian inference is then carried out via integrated nested Laplace approximations (INLA, \citealp{rue2009INLA}).

While this yields an estimate of cortical activation, selection is a post-hoc step based on the excursion set method of \citet{bolin2015excursion}: the joint posterior probability map is evaluated against a specified effect-size threshold, which defines how large an estimated activation must be to count as a true activation, together with a significance level. Evaluating both jointly across all locations avoids the multiple-comparisons problem that would otherwise arise from testing each location separately.

These steps are applied to each subject's individual time-series observations for single-subject modeling. For multi-subject modeling, \citet{Mejia02042020} propose, for computational reasons, to first fit single-subject models and then combine the resulting subject-level posteriors -- either by treating each subject's point estimate as a fixed, known input to a second spatial model, or, at greater computational cost, by propagating each subject's full posterior uncertainty into a joint group-level estimate.

For the data situation handled in this work, no within-subject time-series observations are present, so that single-subject models are neither possible nor necessary. In addition, the condition contrast is modeled rather than the raw measurements. Hence, the modeling approach applied here follows the basic idea of \citet{Mejia02042020} rather than its original intended use case.

Therefore, only a one-stage Bayesian GLM is applied, in which each subject's precomputed condition contrast is treated as one repeated observation of a shared underlying spatial field, and the GMRF prior is fit directly across these observations. This preserves the part of their method that is of interest here -- the spatial smoothing/estimation mechanism itself, i.e., how well a Mat\'ern GMRF prior over the cortical mesh recovers a true spatial field from noisy per-location observations. In particular, the temporal-autocorrelation component normally modeled by the corresponding implementation in the \texttt{R} package \texttt{BayesfMRI} is not applicable here and is disabled.

\subsection{Selection Threshold}

The excursion method for selection requires a specified effect-size threshold $\gamma$ (as \texttt{BayesfMRI} internally requires strictly positive location-wise means, a constant offset is added prior to fitting and subtracted again from all reported estimates and thresholds, including $\gamma$, below). \citet{Mejia02042020} use $\gamma = 0$ in their simulation study and $\gamma \approx 0.027$ -- corresponding to roughly 1\% of the global baseline signal and motivated as a biologically meaningful activation magnitude -- in their empirical application. No comparable percentage-of-baseline threshold is available here, however, since the modeled quantity is a condition contrast rather than a raw signal with an interpretable positive baseline. Another heuristic is therefore required for the present data.

Within the simulation study, four factorial thresholds $\gamma_h \in \{0, 0.2, 0.5, 0.8\}$ are considered, where $\gamma_h = 0$ is an unconstrained baseline matching \citet{Mejia02042020}'s own simulation choice. Each is multiplied by the standard deviation of the measurements, $\gamma = \gamma_h \cdot \mathrm{sd}(\text{measurement})$, computed separately for each hemisphere and simulation replication.

Since the excursion set implementation in \texttt{BayesfMRI} supports only a one-sided exceedance test, and the condition contrasts of interest here may be positive or negative, two one-sided excursion sets are computed per hemisphere (one testing for exceedance above $+\gamma$, one for values below $-\gamma$) and a location is selected if either test is significant. At an overall significance level of $\alpha = 0.05$, each one-sided test is evaluated at $\alpha / 2 = 0.025$, retaining the nominal significance level.

\subsection{Simulation Study Results}

\begin{figure}[!b]
\centering
\includegraphics[width=\textwidth]{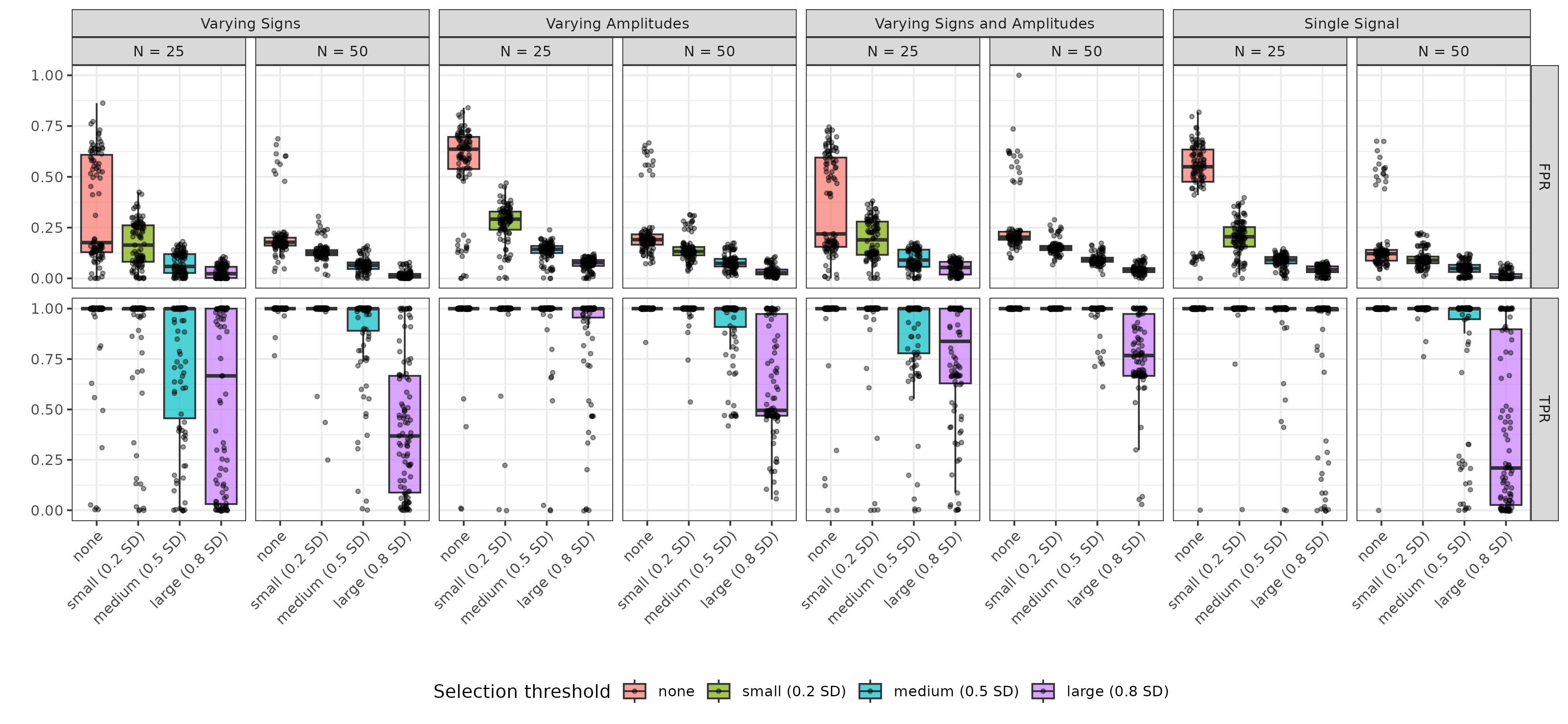}
\caption{Simulation results for FPR (first row) and TPR (second row) of Bayes GLM. Columns correspond to the four simulation settings: varying signs, varying amplitudes, varying signs and amplitudes, and single signal. Thresholds are factors of the standard deviation: $\{0, 0.2, 0.5, 0.8\} \cdot \mathrm{sd}(\text{measurement})$.}
\label{fig:FPR_TPR_rates_thresholds}
\end{figure}

Figure~\ref{fig:FPR_TPR_rates_thresholds} shows the Bayes GLM simulation results, with false positive rates in the top row and true positive rates in the bottom row. The Bayes GLM is fairly sensitive to sample size: with $N = 50$ subjects it shows substantially reduced variance and, in most settings, lower false positive rates than with $N = 25$. With a threshold of $\gamma_h = 0.8$, the method misses a substantial share of signals across all configurations, reflected in true positive rates clearly below 1. Only $\gamma_h = 0$ and $\gamma_h = 0.2$ ensure detection of most signals; however, the varying-signs scenario with $N = 25$ still shows signal-detection problems, with some outlier seeds reaching a true positive rate of 0. The more reliable detection of signals automatically comes with an increased false positive rate across all scenarios.

\begin{figure}[!t]
\centering
\includegraphics[width=\textwidth]{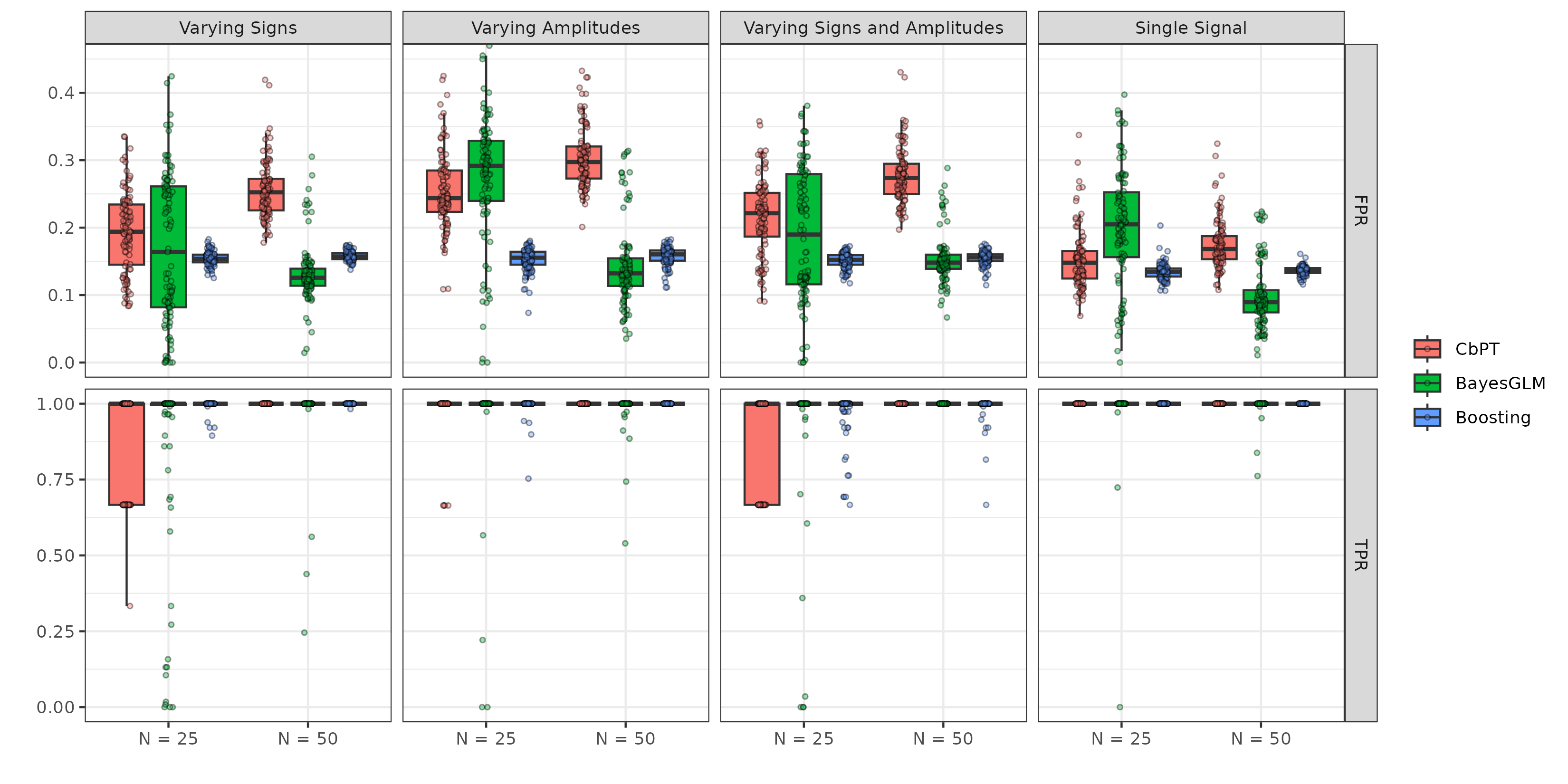}
\caption{Simulation results for FPR (first row) and TPR (second row). The cluster-based permutation test is shown in red (CbPT), Bayes GLM in green (BayesGLM), and Boosting in blue (Boosting). Columns correspond to the four simulation settings: varying signs, varying amplitudes, varying signs and amplitudes, and single signal. Bayes GLM threshold: $0.2 \cdot \mathrm{sd}(\text{measurement})$.}
\label{fig:sim_res_rates_three}
\end{figure}

Since the results on signal detection (true positive rates) barely differ between $\gamma_h = 0$ and $\gamma_h = 0.2$ but the former comes with highly increased false positive rates, $\gamma_h = 0.2$ is chosen for the methods comparison that follows. Figure~\ref{fig:sim_res_rates_three} compares the false positive and true positive rates of Bayes GLM at $\gamma_h = 0.2$ against cluster-based permutation tests and boosting from the simulation study. The robustness of the Wendland boosting approach stands out most clearly. With more subjects, Bayes GLM on average yields lower false positive rates, though outlier seeds with true positive rates below one remain present across all settings except varying signs and amplitudes with $N = 50$.

The estimation routine of the Bayes GLM does not depend on the threshold $\gamma_h$, so no distinction by threshold is necessary here. Table~\ref{tab:sim_errormetrics_BayesBoost} compares the MAE (top) and RMSE (bottom) of Bayes GLM with the proposed Wendland boosting approach. Across all settings, boosting shows both smaller MAE and RMSE, with smaller standard deviations. This reflects the basis-function-wise selection inherent to boosting: estimates in non-selected regions are set to exactly zero, avoiding the additional estimation error that a fully continuous field -- such as the Bayes GLM posterior mean -- incurs in inactive regions.

\begin{table}[!ht]
\centering
\begin{tabular}{ll|cc}
\toprule
Scenario & N & BayesGLM $\text{mae}_\tau$ ($\pm$ sd) & Boosting $\text{mae}_\tau$ ($\pm$ sd) \\
\midrule
Varying & N = 25 & 0.0347 $\pm$ 0.0054 & 0.0110 $\pm$ 0.0013 \\
Signs & N = 50 & 0.0310 $\pm$ 0.0032 & 0.0111 $\pm$ 0.0009 \\ \hline
Varying & N = 25 & 0.0399 $\pm$ 0.0048 & 0.0112 $\pm$ 0.0010 \\
Amplitudes & N = 50 & 0.0357 $\pm$ 0.0037 & 0.0113 $\pm$ 0.0009 \\ \hline
Varying Signs & N = 25 & 0.0394 $\pm$ 0.0049 & 0.0133 $\pm$ 0.0013 \\
and Amplitudes & N = 50 & 0.0353 $\pm$ 0.0033 & 0.0132 $\pm$ 0.0009 \\ \hline
Single & N = 25 & 0.0306 $\pm$ 0.0043 & 0.0093 $\pm$ 0.0014 \\
Signal & N = 50 & 0.0261 $\pm$ 0.0032 & 0.0095 $\pm$ 0.0011 \\
\toprule
 & & BayesGLM $\text{rmse}_\tau$ ($\pm$ sd) & Boosting $\text{rmse}_\tau$ ($\pm$ sd) \\
\midrule
Varying & N = 25 & 0.0501 $\pm$ 0.0065 & 0.0371 $\pm$ 0.0040 \\
Signs & N = 50 & 0.0482 $\pm$ 0.0041 & 0.0370 $\pm$ 0.0027 \\ \hline
Varying & N = 25 & 0.0578 $\pm$ 0.0059 & 0.0379 $\pm$ 0.0030 \\
Amplitudes & N = 50 & 0.0557 $\pm$ 0.0044 & 0.0380 $\pm$ 0.0021 \\ \hline
Varying Signs & N = 25 & 0.0593 $\pm$ 0.0061 & 0.0450 $\pm$ 0.0043 \\
and Amplitudes & N = 50 & 0.0572 $\pm$ 0.0041 & 0.0446 $\pm$ 0.0028 \\ \hline
Single & N = 25 & 0.0448 $\pm$ 0.0058 & 0.0320 $\pm$ 0.0049 \\
Signal & N = 50 & 0.0420 $\pm$ 0.0043 & 0.0326 $\pm$ 0.0036 \\
\bottomrule
\end{tabular}
\caption{MAE (top) and RMSE (bottom) for Bayes GLM, computed from the posterior mean field estimate, compared against the Wendland Boosting approach.}
\label{tab:sim_errormetrics_BayesBoost}
\end{table}

\subsection{Application to the MEG Jaw Experiment}

\begin{figure}[!h]
\centering
\begin{minipage}{0.48\textwidth}
  \centering
  \includegraphics[trim={5cm 3cm 1cm 0.3cm},clip,width=\textwidth]{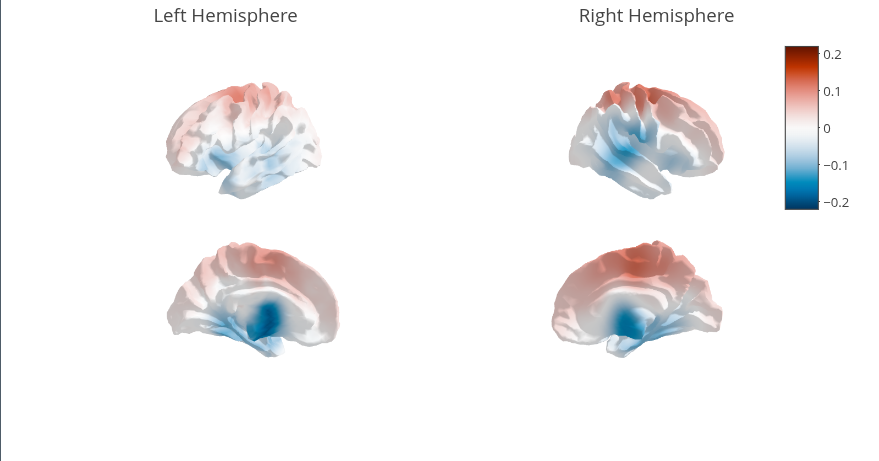}
\end{minipage}\hfill
\begin{minipage}{0.48\textwidth}
  \centering
  \includegraphics[trim={5cm 3cm 2.5cm 0.3cm},clip,width=\textwidth]{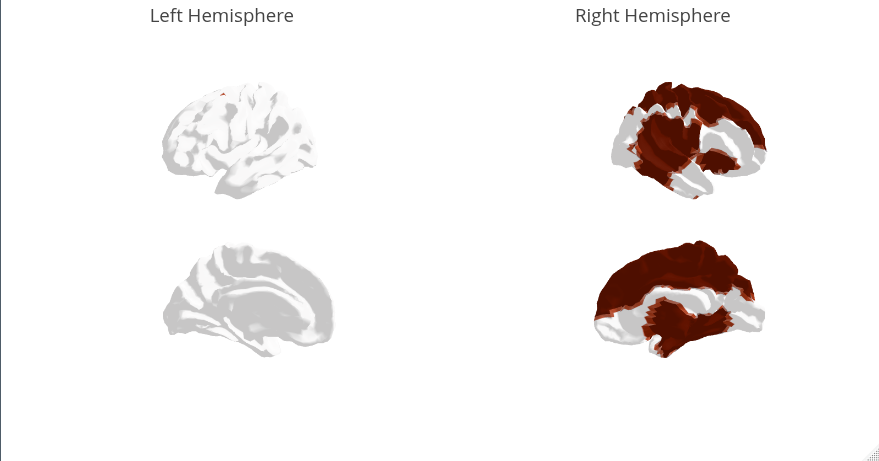}
\end{minipage}
\caption{Empirical results from Bayes GLM with threshold $0.2 \cdot \mathrm{sd}(\text{measurement})$, computed separately per hemisphere. Left panel: effect estimation; right panel: selected regions.}
\label{fig:data_BayesGLM}
\end{figure}

Figure~\ref{fig:data_BayesGLM} illustrates the application of the Bayes GLM to the MEG jaw experiment, with estimation results in the left panel and the selected regions in the right panel. Since the point estimate does not come with an inherent selection mechanism, the (unthresholded) estimates shown in the left panel essentially depict a spatially smoothed version of the averaged observed hemisphere measurements. The post-hoc excursion-based selection step detects a signal existing of one measurement location in the left hemisphere, which is in the superior part of the brain with a positive estimate. In the right hemisphere, Bayes GLM detects signals in both the superior and inferior regions. Of the 2314 vertices selected in the right hemisphere, 1191 show a positive (superior parts) and 1123 (inferior parts) a negative estimated contrast.

